\documentclass[12pt,preprint]{aastex}

\def\kms{km ${\rm s}^{-1}$}
\def\sn1{${\rm s}^{-1}$}
\def\cmd{${\rm cm}^{-3}$}
\def\cmn{${\rm cm}^{-2}$}
\def\ergs{ergs ${\rm s}^{-1}$}
\def\xiu{${\rm erg~cm/s}$}
\def\ngc{NGC~3783}
\def\apm{APM~08279+5255}

\shorttitle{Ionization structure and kinematics}
\shortauthors{Ramirez}

\begin{document}

\bibliographystyle{apjsty}

\title{Kinematics from spectral lines for
AGN outflows based on time-independent
radiation-driven wind theory}

\author{Jos\'e M Ram{\'i}rez}
\affil{Leibniz-Institut f\"{u}r Astrophysik Potsdam, An der Sternwarte 16, 14482 Potsdam, Germany}
\email{jramirez@aip.de}

\begin{abstract}
We build a bulk velocity-dependent photoionization model of the warm absorber
of the Seyfert 1 galaxy \ngc. By adopting functional forms for velocity of the flow 
and its particle density with radius, appropriate for radiation driven winds, we
compute the ionization, temperature, line Doppler shift, and line optical depths 
as a function of distance. 
By doing this we obtain detailed profiles for the entire absorption line spectrum. 
The model reproduces the observed relationship between the gas ionization and the 
velocity shift of the line centroids as well as the asymmetry of the absorption 
lines in the X-ray spectrum
of NGC 3783. It is found that the distribution of asymmetry
requires the presence of two outflows: a higher ionization component 
responsible for the blue wings of the high ionization lines and the red wings of 
the low ionization oxygen lines, and a lower ionization flow that makes up for the 
blue wings of the oxygen lines. Our model predicts a relationship between the
X-rays and the UV absorbers, where the component creating the long 
wavelength lines in the X-ray spectrum is also responsible for creating at least part of the UV absorption troughs. 
\end{abstract}




\section{Introduction}

X-ray (0.5--10 keV)
{\it Chandra} spectra of type-I active galactic nuclei
(AGN) often show presence of absorption lines coming from
H- and He-like ions of O, Ne, Mg, Si, S as well as
from Fe~{\sc xvii}-Fe~{\sc xxiii} L-shell transitions. 
These lines are near the 
region of the bound-free absorption
edges of O~{\sc vii} and O~{\sc viii} at
$\sim0.8$ keV, which are
the hallmark of 
warm absorbers \citep{george1998a,komossa1999a}.
In Seyfert 1 galaxies these 
absorption lines can be narrow
(Seyfert 1 NALs) with
full width at half maximum (FWHM)
spanning $\sim$ 100--500 \kms \space
\citep{kaspi2000a}, or 
broad 
(Seyfert 1 BALs) with
FWHM $\sim$ 500--2000 \kms \space
\citep{kaspi2001a}.
The troughs of the lines are blue-shifted relative to rest frame of
the host galaxy
by velocities that go from several
hundred to a few thousand \kms \space
\citep{yaqoob2003a}, implying the presence of outflows
that cover a wide range of velocities and
ionization stages
\citep{kaastra2000a,kaspi2002a,krolik2001a,rozanska2006a}.

In Seyfert
galaxies and Quasars absorption spectra show a range of
ionization stages that arise from 
ionization parameters that span $\xi \sim 10-1000$
erg cm \sn1 \citep{kaspi2002a}. 
Simultaneous UV and X-ray
observations may enhance this
range even further \citep{shields1997a,gabel2005a}.
The location of these absorbing material
is uncertain. Models
of X-ray absorbers in AGN place them
in a wide range of distances from the central
source, from
winds originated at the accretion
disk \citep{murray1995a,elvis2000a}, 
out to the dusty
($\sim 1$ pc) torus \citep{krolik2001a}
and beyond the narrow-line
region \citep[e.g.][]{ogle2000a}. 

In the Seyfert galaxy
NGC 3783 the 900 ks {\it Chandra}
spectrum of \citet{kaspi2002a}
allows for precise measurements of
radial velocities and widths of the
lines. It is seen 
that the velocity shift of lines 
from Fe~{\sc xxiii}-Mg~{\sc xii}
cover a range of $\sim 60-600$ \kms \space while 
the lowly ionized lines Si~{\sc xiii}-O~{\sc vii}
cover velocities $\sim 500-1000$ \kms \space
(see Figure 6 in Ramirez et. al. 2005). The
average velocity of the warm absorber outflow
of NGC 3783 is around $\sim 500$ \kms.
The spectrum also reveals that the line profiles are asymmetric
\citep{kaspi2002a}, in such a way that
approximately 90\% of the lines
have extended blue wings.
Such asymmetries were quantified by
Ramirez et al. (2005).
In terms of ionization, mostly high ionization species are seen in the
short-wavelength portion of the spectrum
$\sim 4-12$ ${\rm \AA}$. Here, resonant
lines from
Fe~{\sc xxiii}, Fe~{\sc xxii}, Fe~{\sc xxi},
S~{\sc xvi}, S~{\sc xv}, Si~{\sc xiv},
and Mg~{\sc xii} cover
ionization parameters $\xi$ from $\sim 630$ to $\sim 150$
ergs cm ${\rm s}^{-1}$. 
The longer wavelength of the spectrum, 
$\sim 10-20$ ${\rm \AA}$, is dominated by
lower ionization lines from Si~{\sc xiii},
Fe~{\sc xviii}, Ne~{\sc x}, Mg~{\sc xi}, Fe~{\sc xvii},
Ne~{\sc ix}, O~{\sc viii}, and O~{\sc vii} that span ionization parameters
$\xi$ from $\sim 125$ ergs cm ${\rm s}^{-1}$ down to $\sim 8$
ergs cm ${\rm s}^{-1}$. 

Thus, photoionization modeling 
shows that the observed spectrum 
is hardly explained by
a single ionization parameter.
Rather two or three components are needed, and this has been the subject of
well-detailed studies
\citep{krongold2003a,netzer2003a,krongold2005a}.
In those works two/three components in pressure equilibrium
are enough to account for all the charge states seen in the
spectra. When start runing those simulations, the ionization parameter
is not constant throughout the cloud but it varies
in a self consistent way.
On the other hand \citet{steenbrugge2003a}
used a
nearly
continuous distribution of ionization parameters that span three orders of
magnitude.
In reality both approaches do not need to be necessarily contradictory.
We just have to look at the possibility of introducing species between
the components proposed by
\citet{krongold2003a,netzer2003a,krongold2005a};
and that immediately lead us to the discussion of
whether the material absorbing is clumpy or not.
And that is in part the goal of the present work.
We present both possibilities, continuous {\it vs} cumply flows,
with their physical implications along the article.

Although there is no evident 
correlation between the ionization of the absorption species and
the velocity shifts of lines \citep{kaspi2002a},
\citet{ramirez2005a} open the possibility
for such relationship to exist,
as the lines are consistent with being originated from an outward 
flowing wind. 
In that work
a wind velocity law was adopted \citep{castor1975a},
with an ionization fraction that goes as a power-law 
of the ionization parameter
$q\sim \xi^{\eta}$ with $\eta>0$ for all
$\xi$ and all lines.
Further, $q\sim \xi^{\eta}$
and an optical depth of
the form proposed by the SEI method
from \citet{lamers1987a} were used to fit
the line in the spectrum and study 
the possible relationship
between the profiles and the expansion velocity of the 
flow.

Here, in this theoretical approach no analytical
dependence of $q$ with $\xi$ is assumed,
but computed self-consistently using the photoionization code
XSTAR with modifications of the optical depth of the lines
for working with expanding outflow.

It is the aim of the present paper
to further study the wind scenario for the
warm absorber of NGC 3783 by modeling a photoionized wind
flow and trying to reproduce the main features in the
spectrum in terms of both equivalent widths and line profiles.
The model consists of a spherically symmetric gas flow with increasing
velocity with radius according to 
the wind velocity law of \citet{castor1975a}. The microphysics of the
plasma is then
solved in detail along the wind, and radiative transfer is treated for
the flowing plasma including Doppler shift effects on the emerging 
spectrum.    
This kind of approach has been used
previously in  studies of the
ionization and thermal properties
of O stars by \citet{drew1989a},
cataclysmic variables by
\citet{drew1985a} and
in an evaluation of
absorption line profiles
from winds in AGN
by \citet{drew1982a}.

Self-consistent hydrodynamic modeling of 
AGNs have been performed in the past \citep[see e.g.][]{murray1995a}
which qualitatively predict the  same 
$\beta$-velocity law,
where $v(r)\sim (1-r_0/r)^{\beta}$, as adopted here.
However, these models did not include detailed treatment of thermal and spectral
processes that lead to synthetic spectra.

The present paper is organized as follows:
In \S \ref{signi1} we discuss the statistical
significance of the X-ray lines under study.
Later in \S \ref{meth}, we present the theoretical method
we have used in this work.
In \S \ref{results1} we present the main results of
our work. We present the final physical
solution in \S \ref{twoouts1}.
And we summarize in \S \ref{summ1}.

\section{Significance of the shifting in the X-ray Lines}
\label{signi1}

\begin{figure}
\resizebox{15cm}{!}{\includegraphics[angle=0]{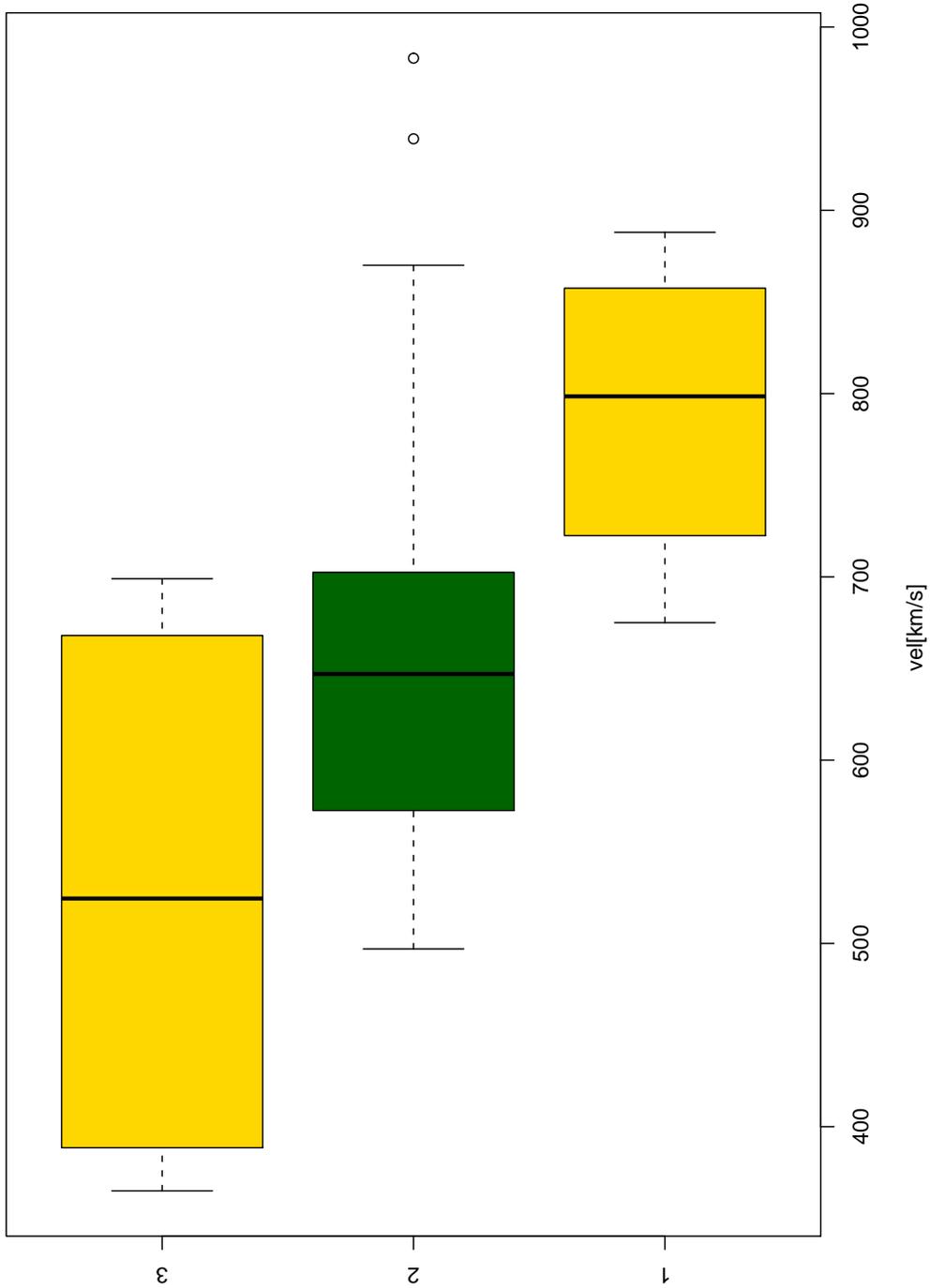}}
\caption{Statistical significance of the shifting of the X-ray
lines found in the spectrum of NGC 3783.
Each box is made of five-number summaries:
the smallest observation
(sample minimum, which are the extreme thin bars at the left of each box),
a lower quartile
(Q1, which is the left thick border of each box), median
(Q2, black thick vertical line),
upper quartile
(Q3, the right thick border of each box) and
largest observation
(sample maximun, which are the extreme thin bars at the right of each box).
\label{boxplot1}}
\end{figure}

Since our analysis searches for possible correlation
between ionization state of the ions forming the absorption
lines observed in the X-ray spectrum of this object,
and the Doppler velocity shift of those lines,
we first of all study the statistical significance
of this possible shifting.
A convenient way to do that is through the
separation of the ions into groups and look for the velocity
shifting.
We take all the ions we find in Table \ref{tblasim1}.
The line's centroid are taken from \citet{kaspi2002a}.
The ionization fraction curves for the classification
are taken from \citet{kallman2001a,bautista2001a}.
Afterwards we grouping them into three groups
\footnote{We group and generate the figure using the
statistical package R - {\tt http://www.r-project.org}}:
Group (1), representative of all ions of low ionization
with ionization parameter $0<\xi({\rm ergs~cm~s^{-1}})<10$;
Group (2) $10<\xi({\rm ergs~cm~s^{-1}}) <250$ representative of intermediate
ionization state; and
Group (3) $250<\xi({\rm ergs~cm~s^{-1}})<650$. A graphical way to represent this
grouping scheme is with a boxplot shown in Fig \ref{boxplot1}.
Each box is made of five-number summaries:
the smallest observation 
(sample minimum, which are the extreme thin bars at the left of each box),
a lower quartile 
(Q1, which is the left thick border of each box), 
median
(Q2, black thick vertical line), 
upper quartile 
(Q3, the right thick border of each box) and
largest observation
(sample maximum, which are the extreme thin bars at the right of each box).
The first conclusion is that all the three Q2s are different.
It is clear that Groups 1 and 3 are different at $\gtrsim 1\sigma$.
There is overlap between Groups 1 and 2 
(possible due to limited resolution of the instrument), but Groups 2 and 3
are different at $\gtrsim 1\sigma$. We go ahead in our analysis bearing
in mind these overlaps in groups.

\section{Method}
\label{meth}
Let us consider a radiation 
source with $L\sim 10^{44}-10^{47}$
ergs ${\rm s}^{-1}$, arising from a supermassive ($\sim 10^8
M_{\sun}$)
 black hole
(BH). Material
$\sim$ 0.1--1 pc from the BH only needs
to absorb a small fraction
of this energy to be
accelerated to few thousand
\kms \space in Seyfert galaxies and
up to $0.1-0.2c$ in high redshift
Quasars \citep{arav1994a,ramirez2008a,saez2009a,chartas2009a}. 
By conservation of mass 
the number density of hydrogen
can be written in spherical
symmetry as
\begin{equation}
n_H(r)=
\frac{\dot{M}}{4\pi r^2 v(r)\mu m_H},
\label{den1}
\end{equation}
where $\dot{M}$ is the mass-loss
rate, $v(r)$ is the outflow speed
at radius $r$, $\mu$ is the mean
atomic weight per hydrogen atom
and $m_H$ is the hydrogen mass.

We adopted a velocity law $v(r)$
compatible with the predictions of the
radiatively driven wind theory
\citep{castor1975a}. The velocity law
has two fundamental roles. The first is to shift
the frequency of the absorbing lines according
to the Doppler effect. The second is to dilute
the gas density, affecting radiative transfer
across the gas  and consequently the 
ionization and thermal
state of the gas. The velocity law varies with distance as
\begin{equation}
w(x)=w_0+(1-w_0)\left(1-\frac{1}{x}\right)^\beta,
\label{vel1}
\end{equation}
here $w$ is the velocity normalized to the terminal
velocity of the wind $v_{\infty}$, $w_0$ is the
velocity in the base of the wind and $x$ is the distance
normalized to the radius of the central core $r/r_0$.
The parameter $\beta$ is the quantity governing
the slope of the velocity with the distance and its
{\it ad hoc} value
depends on
the type of radiative force acting on the wind.
Analysis of hot stars suggests that 
$0.5 \lesssim \beta \lesssim 1$
\citep{lamers1987a}. 
The evaluation
of radiation-driven wind in AGNs of
\citet{drew1982a}, and more recent
dynamical calculations  
\citep[e.g,][]{murray1995a,proga2000a},
suggest that in AGNs 
$0.5 \lesssim \beta \lesssim 2$.
We have computed models using velocity laws
$\beta=0.5,1,1.5$ and 2, representing from fast
($\beta=0.5$) to slow winds ($\beta=2$).
We found that fast winds were not able to simultaneously
cover the range in velocity and ionization state observed
in the spectrum of NGC 3783, needed for the goal
of this work, and that the best-fit velocity law was $\beta=2$.
This is why for the rest of this work we use
$\beta=2$.
Then, we rewrite the
number density in terms of the
velocity law as,
\begin{equation}
n_H(x)=n_0x^{-2}w^{-1},
\label{den2}
\end{equation}
where
$n_0=\left(\frac{\dot{M}}{4\pi \mu m_H}
r_0^{-2} v_{\infty}^{-1}\right)$.
The absorbing line frequencies are shifted
according to the
Doppler relation,
\begin{equation}
w=
\frac{c}{v_\infty}
\left(1-\frac{\lambda}{\lambda_0}\right),
\label{dop1}
\end{equation} 
where $\lambda_0$
is rest wavelength.

Our models are based on clouds illuminated
by a point-like X-ray source.
The input parameters are the
source spectrum, the gas 
composition, the gas density
$n_H(x)$, and the outflow
velocity $w(x)$, where $x$
is the position of each slab
inside the cloud normalized
to the radius of the most
exposed face to the source,
$r_0$.
The source spectrum
is described by the spectral luminosity
$L_{\epsilon}=Lf_{\epsilon}$,
where $L$ is the integrated
luminosity from
1 to 1000 Ryd,
and 
$\int_{1}^{1000{\rm Ryd}} f_{\epsilon}
d\epsilon=1$. This spectral function
is taken to be a power law
$f_{\epsilon}\sim \epsilon^{-\alpha}$,
with $\alpha=1$. The gas consists
of the following
elements, H, He,
C, N, O, Ne, Mg, Si, S,
Ar, Ca and Fe. We use 
solar abundances of \citet{grevesse1996a},
in all our models.

Thermal
and statistical equilibrium in our models are
computed with the code XSTAR 
\citep{kallman2001a,bautista2001a}. The code includes
all relevant atomic processes and computes
the equilibrium temperature and optical depths 
of the most prominent X-ray and UV lines
identified in AGN spectra. 

We consider two types of models, 
the single absorber model (SA)
and the multicomponent model (MC). In both cases the 
absorption line profiles depend simultaneously 
upon the ionization and the kinematics of the absorbing gas.

\subsection{The Single absorber Model}
\label{sm}

This model consists of a single extended cloud directly in the 
line of sight between the observer and the central source.
Such a cloud flows away from the central source and towards the
observed. For this model 
we make use of the Sobolev
approximation for computing
the
line optical depths
\citep{castor1975a},
\begin{equation}
\tau_{\nu}(r)= \frac{\pi e^2}
{mc}f\lambda_0 n_i(r) \left(\frac{dv}{dr}\right ) ^{-1},
\label{opt1b}
\end{equation}
where $f$ is the absorption oscillator strength,
$\lambda_0$ (in cm) is the laboratory wavelength
of the transition, $n_i$ (in \cmd)
is the number density
of the absorbing ion,
and $dv/dr$ is the velocity gradient in the wind.
This gives us the relation between the
outflow state and the radiation field.
This is different from the calculation
of the optical depth in the static case, 
which is directly proportional to the column density. 

Inside the cloud we use a one-step forward
differencing formula for the radiation
transfer \citep{kallman1982a}
\begin{equation}
L_{\nu}(r+\Delta r)=
L_{\nu}(r)
e^{-\tau_{\nu}(r)}
+4\pi r^2j_{\nu}(r)
\frac{1-e^{-\tau_{\nu}(r)}}
{\kappa_{\nu}(r)},
\label{tra1}
\end{equation}
where $j_{\nu}(r)$ is the
emission coefficient at radius $r$.

Under ionization equilibrium
condition the state of the gas
depends just upon
the shape of the ionizing
spectrum and the ionization parameter $\xi$,
that we define as in 
\citet{tarter1969a}
\begin{equation}
\xi(r)=\frac{4\pi F(r)}{n_H(r)},
\label{xi}
\end{equation}
where $F(r)$ is the total ionizing flux,
 \begin{equation}
 F(r)=\frac{1}{4\pi r^2}
 \int_{1}^{1000 {\rm Ryd}} L_{\nu}(r)d\nu,
 \end{equation}
with $L_{\nu}(r)$ given
by equation (\ref{tra1}).
The requirement that $\xi$ spans various orders of magnitude,
as observed, yields
a cloud geometrically thick throughout most of
the spectrum, i.e. 
a {\it photoionization bounded} cloud.
For instance,  
a luminosity $L\sim 10^{44}$ \ergs and $\Delta \xi = \xi_2-\xi_1=\frac{L}{n}
(\frac{1}{r^2 _2} -\frac{1}{r^2 _1})=10^3$ yield
$n\sim10^{6}$ \cmd \space
and a
column density
of the absorbing material is $N_H\sim n \Delta R \sim 9.7 \times 10^{24}$ cm$^{-2}$.  
This is a large value, such that even lines
with moderately small oscillator strengths
become saturated in the emergent spectrum,
unless the metal abundances are reduced
by several orders of magnitude with
respect to solar.

\begin{figure}
\resizebox{15cm}{!}{\includegraphics[angle=0]{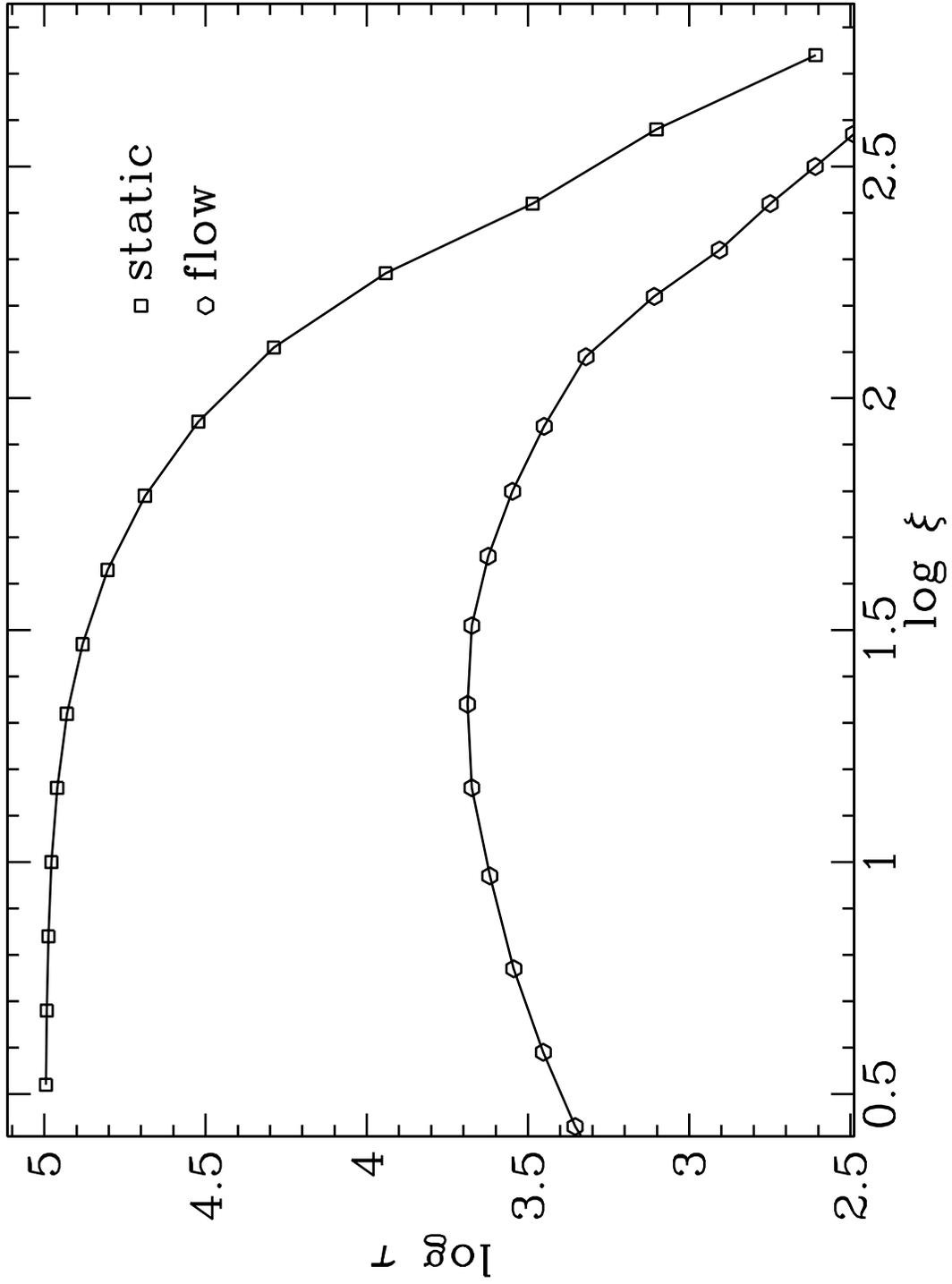}}
\caption{Optical depth ($\tau$) static case {\it vs}
outflow model
for the line Ne~{\sc x} $\lambda$12.134 (see text).
\label{nex1}}
\end{figure}

\begin{figure}
\resizebox{15cm}{!}{\includegraphics[angle=0]{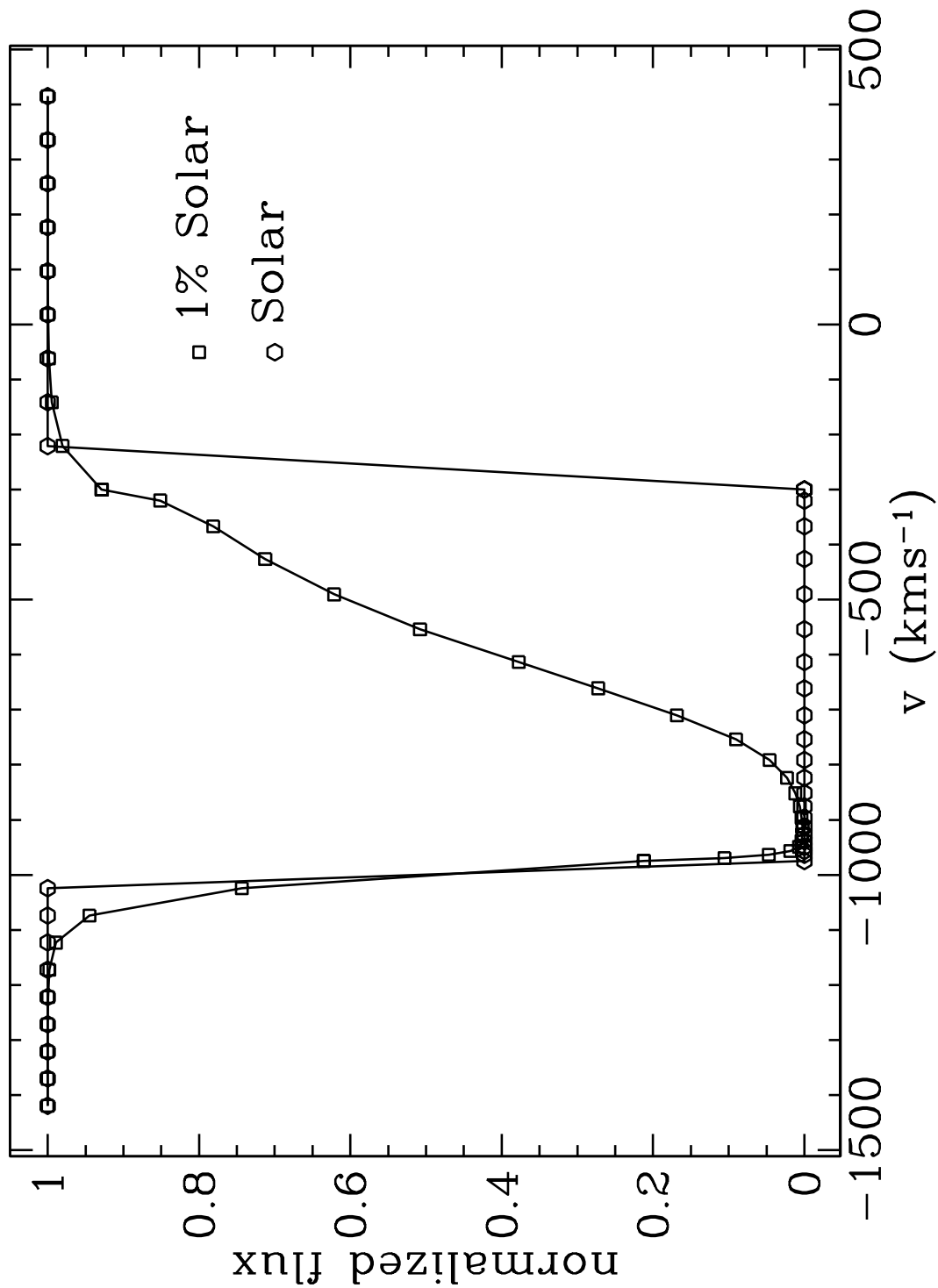}}
\caption{
Profile of the line Ne~{\sc x}~$\lambda 12.134$ formed
by taking $F_{\lambda}(v)=F_c\times \exp[-\tau_{\lambda} (v)]$ as
function of the velocity. The rectangle profile is made by a cloud
with solar abundances (open circles). The profile with an extended
red wing and a sharp blue wing is made by a cloud with 1\% solar
abundances (open square).
\label{nex2}}
\end{figure}

In Figure \ref{nex1}
we compare the computed optical depth
for the Ne~{\sc x} $\lambda$12.134 line
in a stationary nebula with $L=10^{44}$ \ergs, and
constant density $n= 10^{6}$ \cmd \space 
with a flow with 
$n(x)=3.3\times 10^{6}x^{-0.5}w^{-1}$,
a wind velocity function with $\beta$= 2, and $v_{\infty}=1000$
\kms. In both cases we adopt a turbulence velocity
of 200 \kms \space 
and solar abundances. 
In the stationary case, $\tau$ is
proportional the column density of the absorbing
material and can reach very large values.
In the outflow model, $\tau$ peaks at $\log \xi
\sim 1.3$ ($v\sim 890$ \kms) reaching $\sim 5000$,
that is nearly two orders of magnitude smaller than the stationary
cloud for the same $\xi$. 
Such high optical
depths are common to other lines in the X-ray
band, like O~{\sc viii} in the 14-20 ${\rm \AA}$ and  
the line O~{\sc viii} $\lambda 18.969$.
In such cases lines appear saturated in the absorption spectrum
in contrast with observation. In  
\citet{ramirez2005a} we fit the integrated
optical depths of resonant lines in the 
spectrum of NGC 3783  and found
$T_{tot}\lesssim 1$ for most of the lines. 
We illustrate in Figure \ref{nex2}
the consequences of taking this
model for the reproduction of the profile of the line
Ne~{\sc x} $\lambda$12.134. In order to have a non-saturated line (as observed),
we had to reduce the abundance to $\sim 1$ \% solar,
which has no physical motivation. We do not go further with this model,
and present the clumpy (multicomponent) scenario
in the next section.

\subsection{The Multicomponent Model}
\label{mc}

Now, we examine the scenario in which 
the absorption profile seen in the X-ray and UV
spectra of AGNs are made up of multiple
components. The main difference between this model
calculations and previous ones by other authors 
\citep[e.g.,][]{kaspi2002a,krongold2003a,rozanska2006a},
is that in our
model all the components
are linked by a velocity law and a gas density distribution.

Each absorber is specified by an ionization
parameter $\xi$, a column density $N_H$, an absorption
covering factor, a gas density $n_H(r)$ and an
outflow velocity $v(r)$ in order to shift the absorption lines
according
to the Doppler effect. Once the ionic fraction is calculated
from the ionization equations, and the ionic levels
computed, the opacity
in each frequency bin is
\begin{equation}
\kappa_{\nu}\rho=
\frac{\pi e^2}{m_ec}f_l n_l
\left[ 1- \frac{n_u g_l}{n_l g_u}
\right ] \phi (\Delta \nu),
\label{k1}
\end{equation}
where $\kappa_{\nu}$ is the opacity at the frequency
$\nu$, $\rho=\mu m_H n_H$ is the mass density,
$f_l$ is the absorption oscillator strength, 
$n_l$, $n_u$ (in $cm^{-3}$),
$g_l$, $g_u$ are the number density and the statistical weight
of the lower and upper levels of the transition respectively.
We allow for the lines to have a finite width
characterized by the line profile $\phi (\Delta \nu)$,
with a width which is the greater between the thermal and
the turbulent motions. In all our models the turbulence velocity
is assumed to be 200 \kms. The optical depth of a line in
each component is
\begin{equation}
\tau_{\nu}=\int_{r_1}^{r_2}
\kappa_{\nu}(r) \rho(r) dr,
\end{equation}
where 
$r_2$ and $r_1$ are the limits of the cloud.

It is important to highlight
a special difference
between the way we use the ionization
parameter in the MC model and the way it is
used in the SA model. Because of the
clouds intervening in this model
are optically thin, the
ionization parameter at the most exposed
face of the cloud remains essentially
constant through the cloud, i.e. 
\begin{equation}
\xi(r)=\frac{4\pi F(r)}{n_H(r)},
\label{xi1}
\end{equation}
where $F(r)$ is the total ionizing
flux at the radius $r$.

When two or more outflows are put together
they are assumed to be distributed
such that the observed spectrum is the result
of the addition of all components.
So the radiative flux in each bin of frequency
for the composite spectrum is
\begin{equation}
F_{\nu}=\sum_{i=1}^{m} 
f_{\nu}(x_i)
\label{tautot123}
\end{equation}
where $m$ is the number of
absorbing clouds 
in the line-of-sight,
$f_{\nu}(x_i)$ is the flux resulting
from the pass of the continuum
radiation through the absorbing
cloud ``$i$",
and
$x_i$ is the normalized spatial
radius of the cloud.

In \citet{ramirez2005a} is
showed that in order to fit the line profiles
in NGC~3783 a geometry different from the
spherical for the gas distribution is required. Such
deviation from spherical geometry has two effects. In
the normalized notation the number density is
\begin{equation}
n_H(x)=n_0
x^{-2+\kappa}w^{-1},
\label{den3}
\end{equation}
where $0 \leq \kappa \leq 2$.
A positive value of $\kappa$ implies that the
gas flow dilutes more slowly than in a free
spherical expansion, i.e. that there are sources of 
gas embedded in the flow, or that the flow is confined. 
A negative value corresponds to sinks of gas in the flow,
or expansion of an initially confined flow in a flaring 
geometry.
Secondly, we  allow the radiation flux to have a form
\begin{equation}
F=F_0
\times x^{-2-p},
\label{f}
\end{equation}
where $p$ is an index to mimic a deviation
of flux from the pure geometrical dilution case.

This is expected if 
the medium between clouds
has a significant optical depth 
(if $p$ is positive) or
if there are sources of radiation 
embedded in the flow (if $p$ is negative).  


\section{Assumption about the number density}
\label{assumden1}
Before going into the presentation of the results
we would like to highlight important differences
between the underlying physical/geometrical
motivation of the present work and previous
ones.
After reviewing single (geometrically thick model)
and clumpy (geometrically thin model),
we favored the latest one and require for the
production of the line profile found in the spectrum
of Seyfert galaxies, $\Delta R/R << 1$
(also based on results by \citet{gabel2005a} from UV data).

Based on UV CIII$^{*}$ density constraints
\citep{gabel2005a},
the electron density for the absorber could be
$n_e\approx 10^4$ \cmd.
With this (using eq. \ref{xi1} and $n_H\approx n_e/1.2$),
the distance between the absorber and the central source is 
$R\approx 25$ pc ($\approx 10^{20}$ cm).
As in \citet{gabel2005a}, the low ionization species
(XLI in that paper) seems to share the same kinematics
with the UV absorber.
Using typical luminosity $L$ for this object of
$\approx 10^{44}$ \ergs, $n_H \approx 10^4$ \cmd,
ionization parameter of $\xi \approx 1$,
$R \approx 10^{20}$ cm;
similar to that computed by \citet{gabel2005a},
but if we take
$n_H \approx 10^{11}$ \cmd, as is required for $\Delta R << R$;
then $R \approx 10^{16.5}$ cm.
This is not in contradiction with
\citet{krongold2005a}, where they set
limits on the density and location of the absorber;
with $n_e > 10^{4}$ \cmd, and
$D < 5.7$ pc.
However is clearly different from the computation
made by
\citet{netzer2003a},
of
$n_e < 5 \times 10^{4}$ \cmd \space
$D > 3.2$ pc for the $log(U_{ox})=-2.4$ component,
$n_e < 10^{5}$ \cmd \space
$D > 0.63$ pc for the $log(U_{ox})=-1.2$ component,
and
$n_e < 2.5 \times 10^{5}$ \cmd \space
$D > 0.18$ pc for the $log(U_{ox})=-0.6$ component.
It is clear that some of the differences
in the estimations can be due to differences
on the nature of the physics behind the estimations.
In \citet{krongold2005a} (and also \citet{reeves2004a} for the Fe K shell),
the fundamental assumption behind the computation of
$n_e$ is that the absorber responds instantaneously to changes of
flux in the ionizing source, while
in
\citet{netzer2003a} the estimations are based on
temperature derived models,
average recombination rates and no
response to continuum variations on timescale of 10 days.
And, for our purposes, that basically translate in differences between
the two proposed physical mechanisms; thermally accelerated winds,
which from grounds has to consider sublimination radius
for the material not being evaporated,
and radiatively accelerated wind
with origin possibly in the accretion disk
with subparsec scales;
both competitor theories in the explanation of the origen of the
warm-absorber outflows.


\section{Results and Discussions}
\label{results1}

\begin{figure}
\resizebox{15cm}{!}{\includegraphics[angle=0]{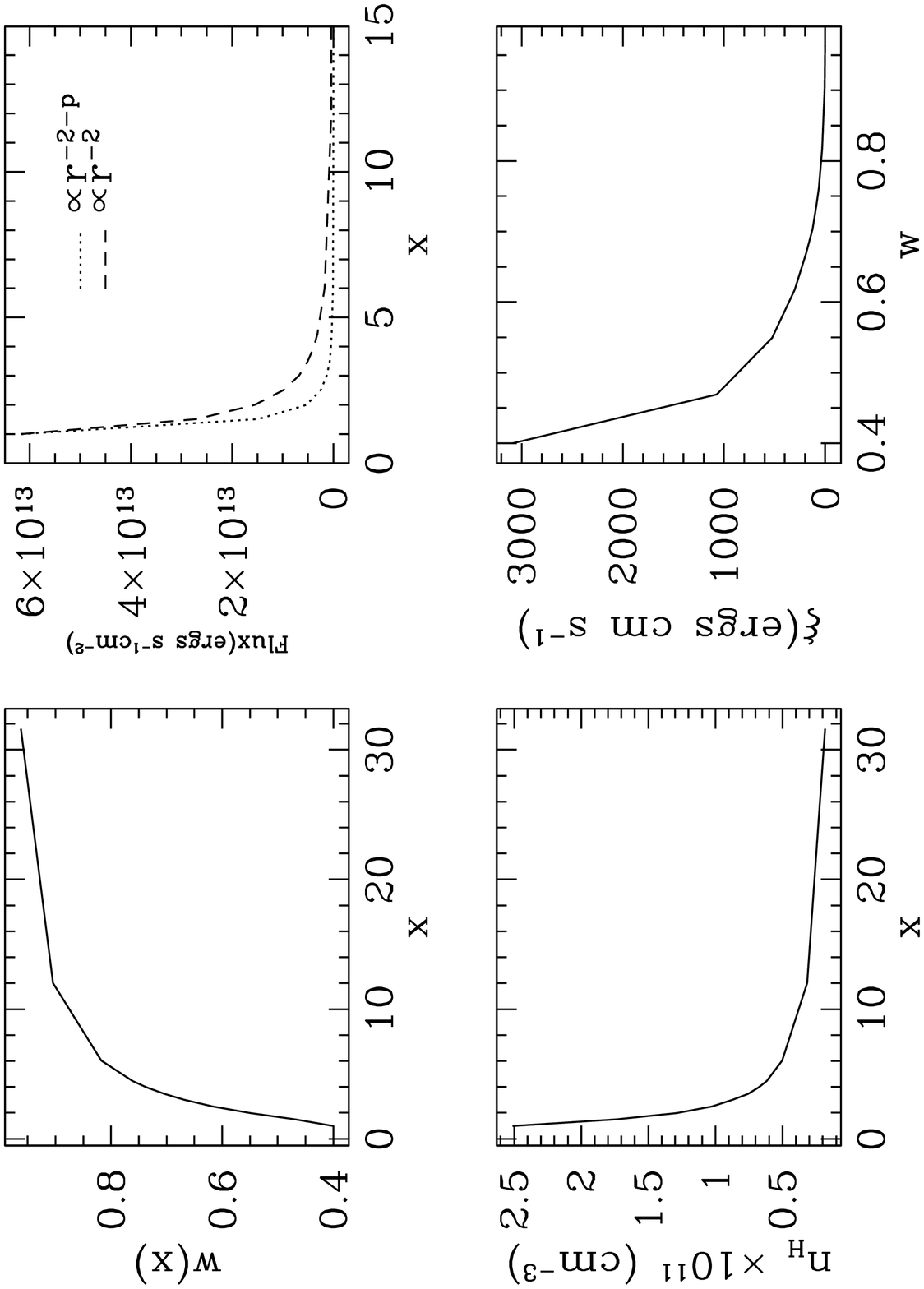}}
\caption{Variation of the main kinematic and ionization variables 
for model A (see text). 
{\it Upper left}: Velocity law $w(x)$ for the model.
{\it Upper right}: The ionizing 
radiative flux, decaying  with $\propto r^{-2}$ (dashed line) and
with $\propto r^{-2-p}$ (dotted line).
{\it Lower left}: Variation of the density as 
function of $x$.
{\it Lower right}: Variation of the ionization parameter ($\xi$) with
the normalized velocity.
\label{fmodelA}}
\end{figure}

\begin{figure}
\resizebox{15cm}{!}{\includegraphics[angle=0]{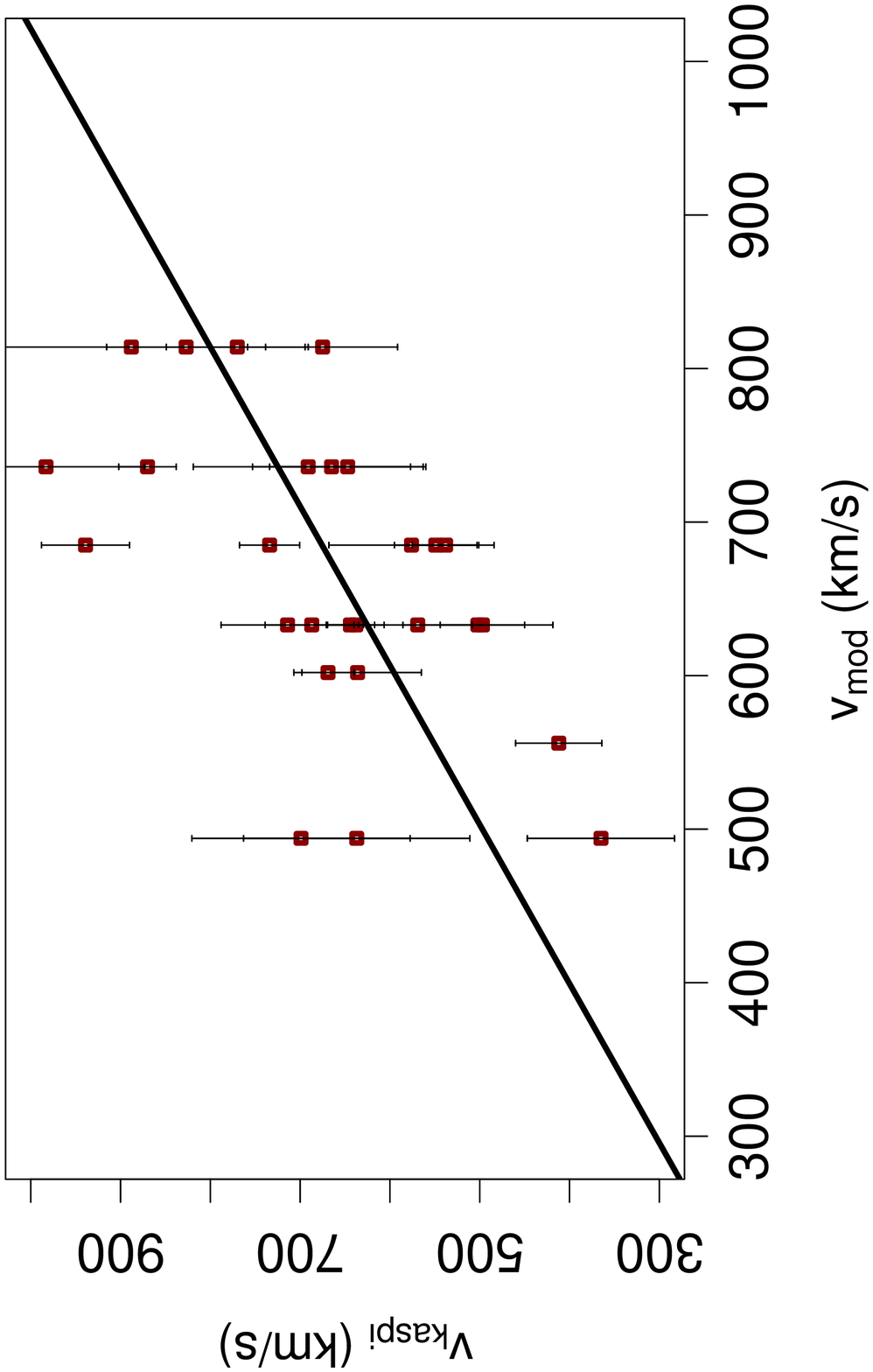}}
\caption{Measured {\it vs} model predicted velocities for
the group of unblended lines given in Table 2 of \citet{ramirez2005a}.
The measurements are taken from \citet{kaspi2002a}.
The solid line is the best line after a linear regression for model A.
\label{mvelA}}
\end{figure}

In Figure \ref{fmodelA} we show the variation
of the different variables governing
the kinematics and the
ionization structure of one of our
models (model~A). 
This is a wind with a velocity slope
$\beta=2$, $\kappa=1.5$ and $p=1.5$.
The number of kinematic component $m$ is
11. The assumed abundance is solar as is given
in Table \ref{tbl1solar}, and each component
has a column density of $N_H=5\times 10^{20}$ \cmn.
In this model the variation in the ionization
parameter is $\log \xi=3.5-(-0.65)$ [\xiu],
where the ``launching" ionization parameter is defined as
$\log \xi_0=3.5$ [\xiu],
and the variation in
density $\log n_H=11.4-10.26$ [\cmd].
We can see the variation with distance of the
input parameters
$w$,
flux,
$n_H$ and
$\xi$,
for model~A in Figure \ref{fmodelA}.
In Figure \ref{mvelA} we plot the  
velocity shifts taken from the maximum 
line optical depths $\tau_{max}$ of
our model versus the line centroids measured by
\citet{kaspi2002a} for NGC 3783 (see Table 3 in that paper).

We fit a linear model by robust regression using the M estimator
(Huber 1964). The weights are included as the
inverse of the variances,
with the variance equal to $\Delta v=(v_{lo}+v_{hi})/2$,
and $v_{lo}$, and $v_{hi}$ are the differences
between centroid and
lower and upper velocity limits
of the measured lines given in Table 3 of \citet{kaspi2002a}.
The best-fit slope
is 0.97$\pm$0.31 (solid line in the figure),
and the residual standard error (rms) 
is 15.4 for 25 degrees of freedom (dof).
This agreement is encouraging considering 
that no model have been suggested
before to explain the
possible
correlation between the ionization
parameter and the velocity shift seen in this Seyfert galaxy.

Although the velocity shifts measured with respect to line minima 
are well explained by this model
the distribution
(in general)
of line profile shapes
is not in agreement with that 
reported in \citet{ramirez2005a},
for instance.
From that study most
lines have extended blue wings.
On the other hand, 
the model does predict the correct asymmetries 
for a few lines from highly
ionized species, but the profiles of the
lower ionizations lines exhibit more extended red wings,
unlike observations. The discrepancy
between the modeled
profile of the 
O~{\sc viii} $\lambda$18.9 line
and the observed in \ngc, is important.
This is why we created model~B.

\begin{figure}
\resizebox{15cm}{!}{\includegraphics[angle=0]{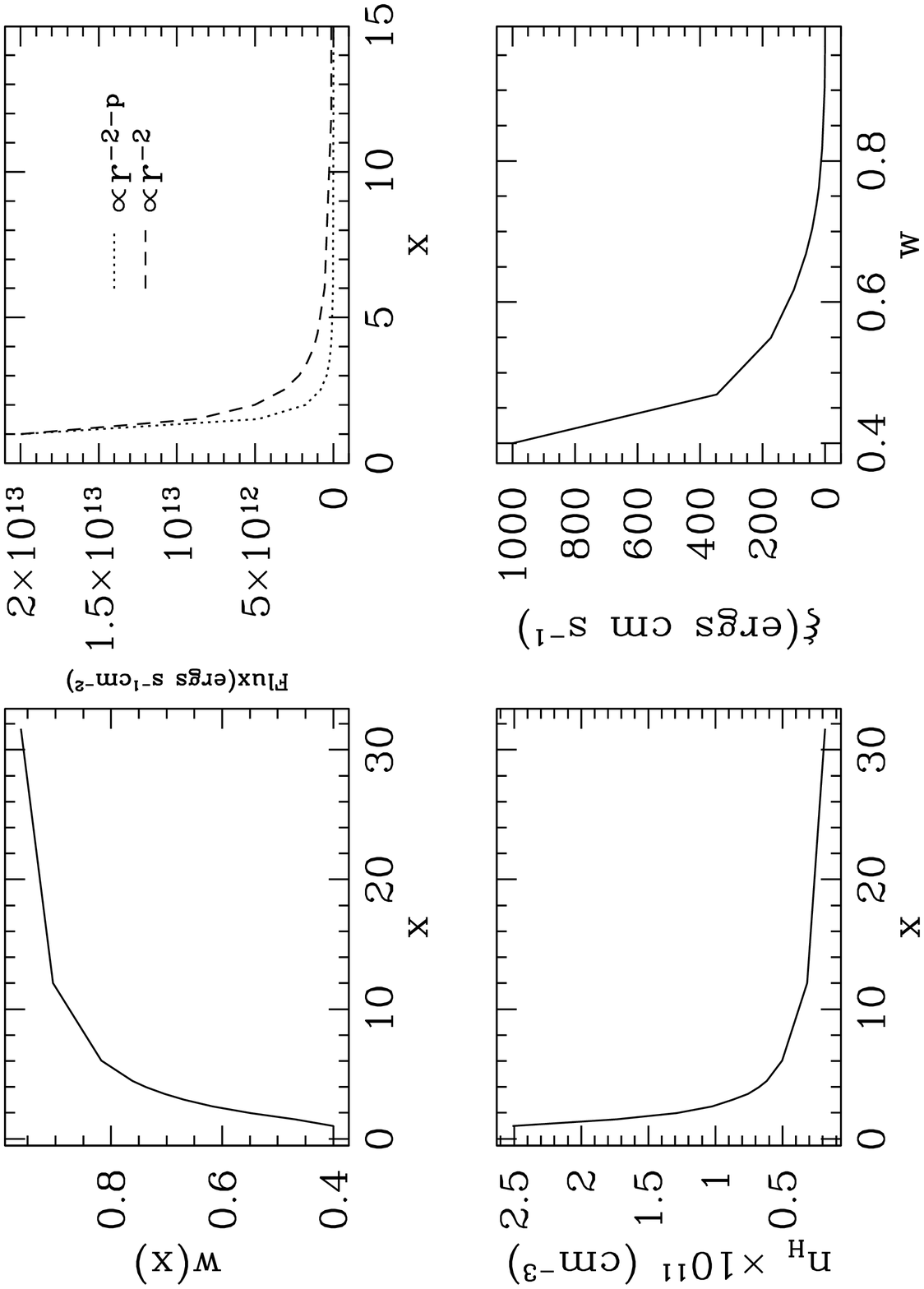}}
\caption{Variation of the main kinematic and ionization variables 
for model B (see text). 
{\it Upper left}: Velocity law $w(x)$ for the model.
{\it Upper right}: The ionizing 
radiative flux, decaying  with $\propto r^{-2}$ (dashed line) and
with $\propto r^{-2-p}$ (dotted line).
{\it Lower left}: Variation of the density as 
function of $x$.
{\it Lower right}: Variation of the ionization parameter ($\xi$) with
the normalized velocity.
\label{fmodelB}}
\end{figure}

\begin{figure}
\resizebox{15cm}{!}{\includegraphics[angle=0]{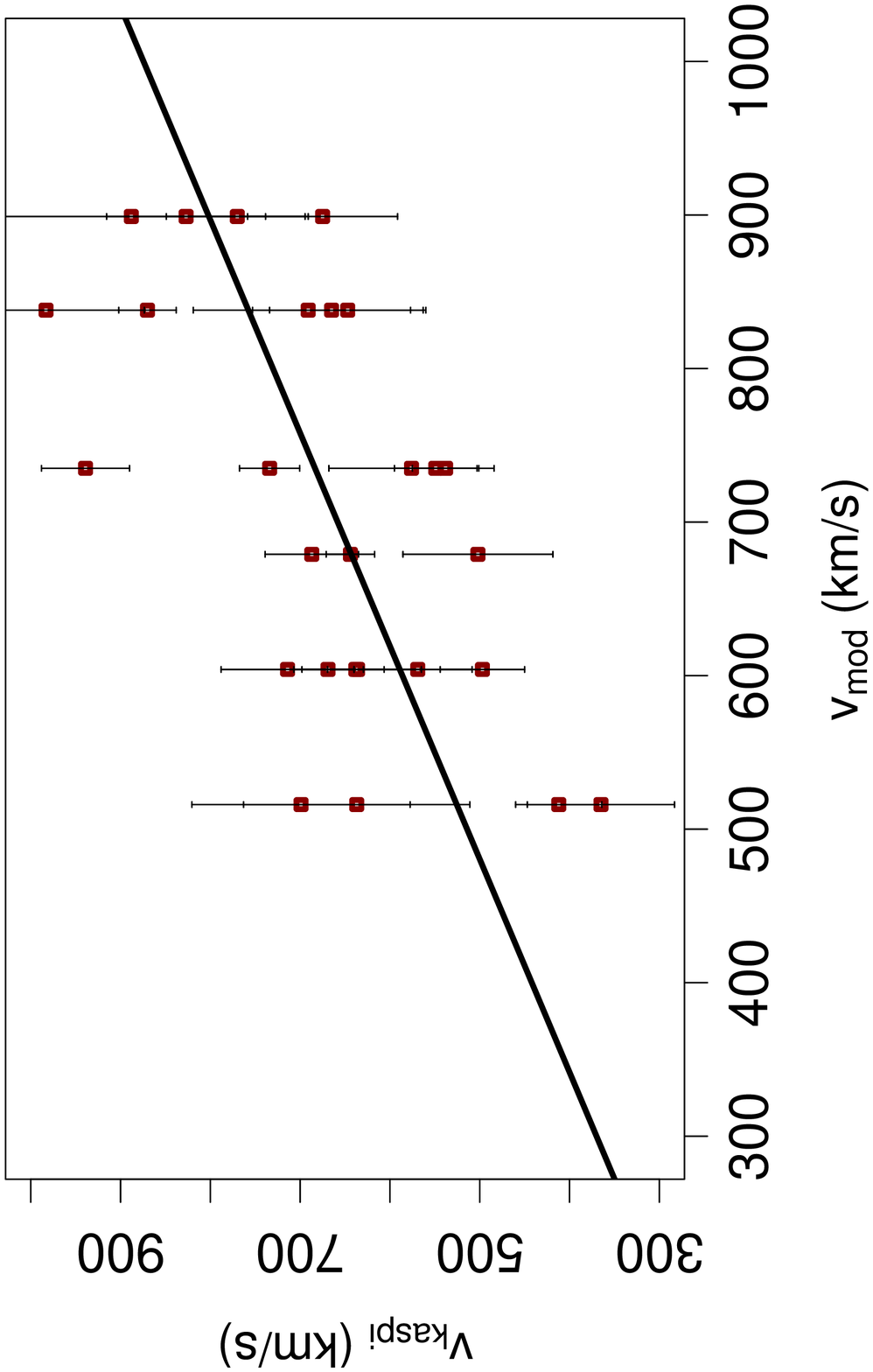}}
\caption{Measured {\it vs} model predicted velocities for
the group of unblended lines given in Table 2 of \citet{ramirez2005a}.
The measurements are taken from \citet{kaspi2002a}.
The solid line is the best line after a linear regression for model B.
\label{mvelB}}
\end{figure}

In Figure \ref{fmodelB} we present 
the variation
of the  variables governing
the kinematics and the
ionization structure of another
model (model B). Here we change slightly
the parameters $\log \xi_0$, and $v_{\infty}$
(see Table \ref{tbl2solar} for details).
The wind has the parameters 
$\beta=2$, $\kappa=1.5$ and $p=1.5$.
The number of kinematic component $m$ is
11. The abundances and the column density are as in model A.
In this model the variations in ionization
parameter and density are $\log \xi=3.00-(-1.13)$ [\xiu]
and 
$\log n_H=11.40-10.26$ [\cmd].
We compare models A and B
in terms of 
the variation of optical depth vs. velocity of the
flows of three important lines, i.e,
Si~{\sc xiv}~$\lambda 6.182$, Si~{\sc xiii}~$\lambda 6.648$
and Mg~{\sc xii}~$\lambda 7.106$.
We could estimate the contribution of each
cloud to the formation of the composed profiles of these lines. 
We have that model~B is superior to model~A in describing the
low-velocity portion of the flow reflected by the high
ionization lines seen in NGC 3783 (extended blue wings),
but yields a worse description of the spectrum at higher terminal velocities
responsible for the center of the lines measured by
\citet{kaspi2002a}.
Figure \ref{mvelB} shows the fit of
a linear model by robust regression using the M estimator
for model B.
As in model A,
the weights are included as the
inverse of the variances,
with the variance equal to $\Delta v=(v_{lo}+v_{hi})/2$,
and $v_{lo}$, and $v_{hi}$ are the differences
between centroid and
lower and upper velocity limits
of the measured lines given in Table 3 of \citet{kaspi2002a}.
The best-fit slope
is 0.72$\pm$0.20 (solid line in the figure),
and the residual standard error (rms)
is 17.6 for 25 degrees of freedom (dof).
In general, the scatter get worse for lines with
terminal velocities greater than
1100 \kms.

\subsection{Modeling global properties - $\xi$ {\it vs} velocity}

One of the prime goals of this study is to examine the
relationship between the kinematics and the ionization
structure of the flow.
This is in complement with earlier
works 
\citep{krongold2003a,netzer2003a,krongold2005a}
which follow the algorithm of constructing
a grid of static photoionization models varying in
ionization parameter and column density with the selection
of this quantities which best reproduce the EW observed
in the spectra, in addition paying attention to the
relationship between these physical parameters and
the position in wavelength space of the
absorption troughs.

\citet{krongold2003a} fitted photoionization models to the
900 ks spectrum of NGC 3783 using two phases;
one at high ionization and high temperature
$\log T \sim 5.98$ (HIP) and one at low ionization
and temperature $\log T \sim 4.41$ (LIP), finding  good agreement 
between the modeled equivalent width (EW) and the
measured for a set of absorption lines (see Figure 9 of
\citet{krongold2003a}). They set each of these
phases at a single outflow velocity of $\sim 750$ \kms,
assuming spatial coexistence of the two absorbers.
The approach is similar to that from \citet{netzer2003a},
but in the latest one they use three components in pressure
equilibrium, with two kinematic components each.

\begin{figure}
\resizebox{15cm}{!}{\includegraphics[angle=0]{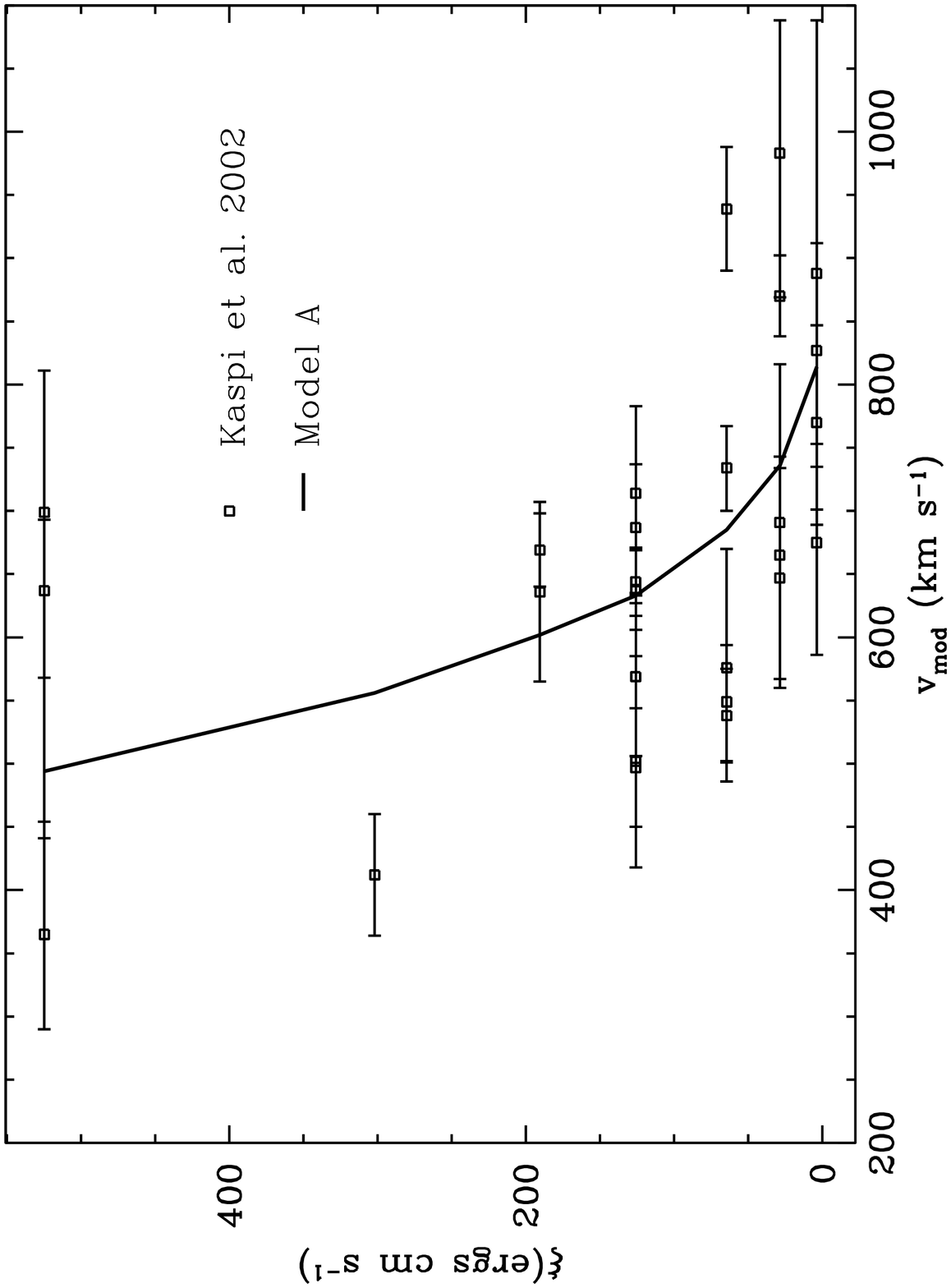}}
\caption{Relationship between the ionization parameter and velocity.
The solid line is the theoretical prediction for model A. The open
squares with error bars are points with velocities taken from
\citet{kaspi2002a}, for the group of lines compared in 
Figure \ref{mvelA}.
\label{mvelxiA}}
\end{figure}

\begin{figure}
\resizebox{15cm}{!}{\includegraphics[angle=0]{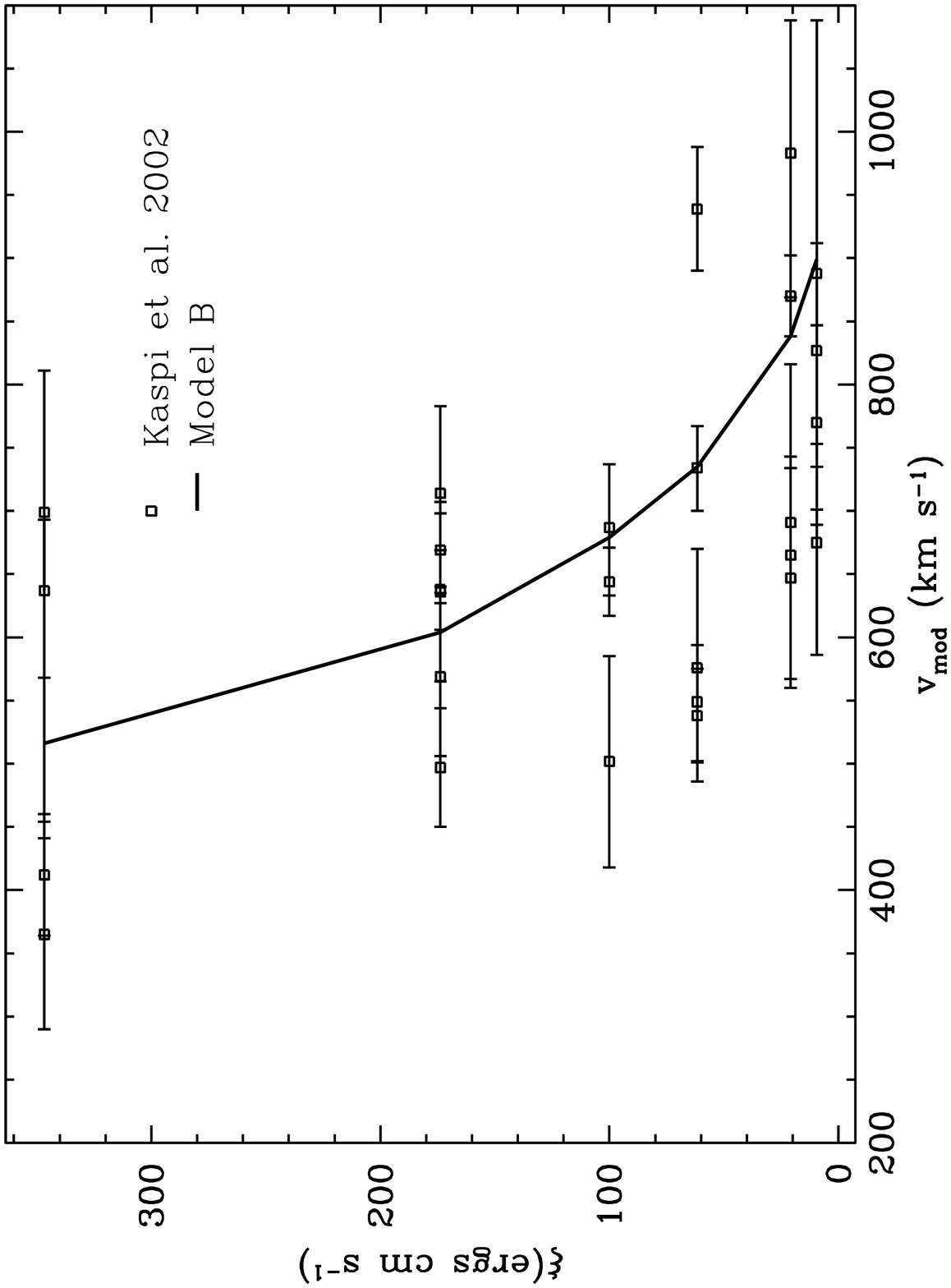}}
\caption{Relationship between the ionization parameter and velocity.
The solid line is the theoretical prediction for model B. The open
squares with error bars are points with velocities taken from
\citet{kaspi2002a}, for the group of lines compared in 
Figure \ref{mvelB}.
\label{mvelxiB}}
\end{figure}

One property of our model is the
capability of open the possibility
of establishing
\footnote{We explicitly state {\it open the possibility}, 
since the limited resolution of the
telescope does not allow us to go beyond this possibility.}
a relationship between the ionization
and the kinematics of the gas based on a radiatively
accelerated wind. Figures \ref{mvelxiA} and
\ref{mvelxiB} show this relation for models A and B
respectively.
In these plots show the predicted relationship between observed
velocities and ionization parameter.
While absorption from highly ionized
ions originates from low-to-intermediate (200-600 \kms)
velocities, lines from lower ionization
stages are formed at intermediate-to-high velocities
(600-1000 \kms).

\subsection{Far-UV and UV absorbers}
\begin{figure}
\resizebox{15cm}{!}{\includegraphics[angle=0]{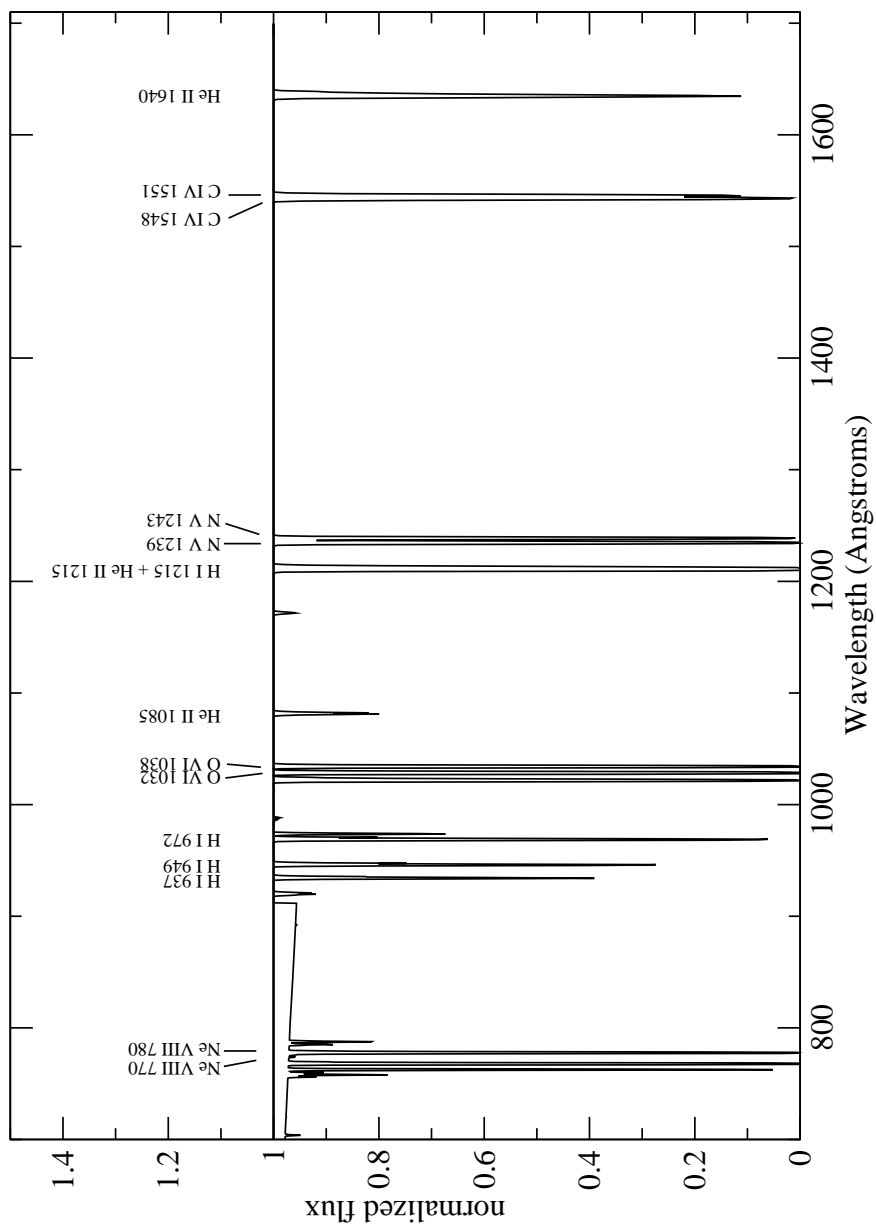}}
\caption{Predicted UV spectrum by model B (in the range 700-1700 ${\rm \AA}$). 
These absorption lines
are formed by the faster (or at lower ionization parameters) clouds
of model B, i.e., $\log \xi =0.12, -0.65$. Most of these lines
are found in real UV spectrum of AGN.
\label{uv700-1700}}
\end{figure}

\begin{figure}
\resizebox{15cm}{!}{\includegraphics[angle=0]{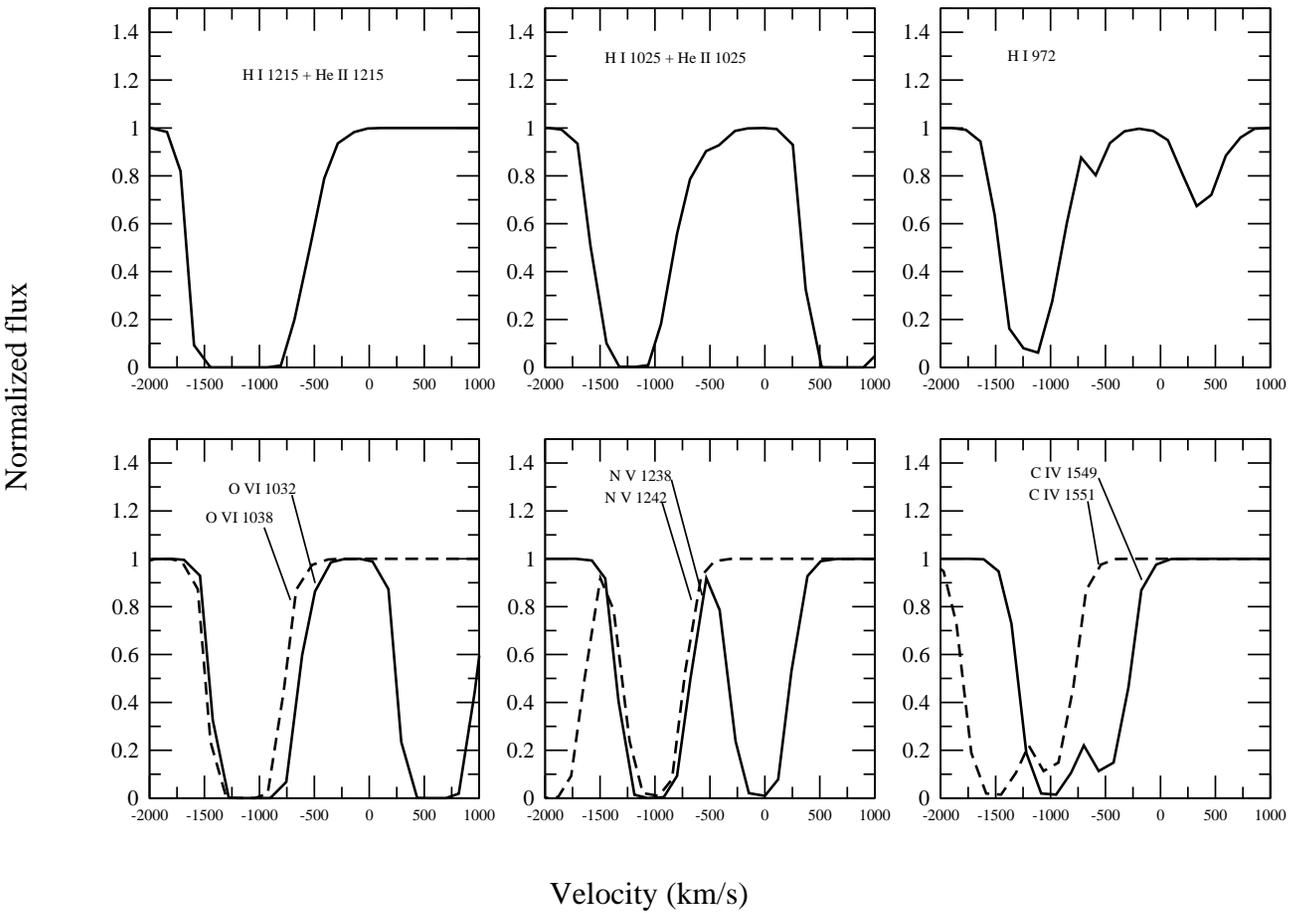}}
\caption{Predicted UV spectrum by model B (in the velocity space). 
These absorption lines
are formed by the faster (or at lower ionization parameters) clouds
of model B, i.e., $\log \xi =0.12, -0.65$. Some of the velocities
displayed by these lines are shared by lines coming from the
X-ray band. For transforming to velocity space, we have taken the shorter
rest-wavelength of the doublet (solid line), and also the
longer rest-wavelength of the doublet (dashed line).
\label{uv-velocity}}
\end{figure}

\begin{figure}
\resizebox{15cm}{!}{\includegraphics[angle=0]{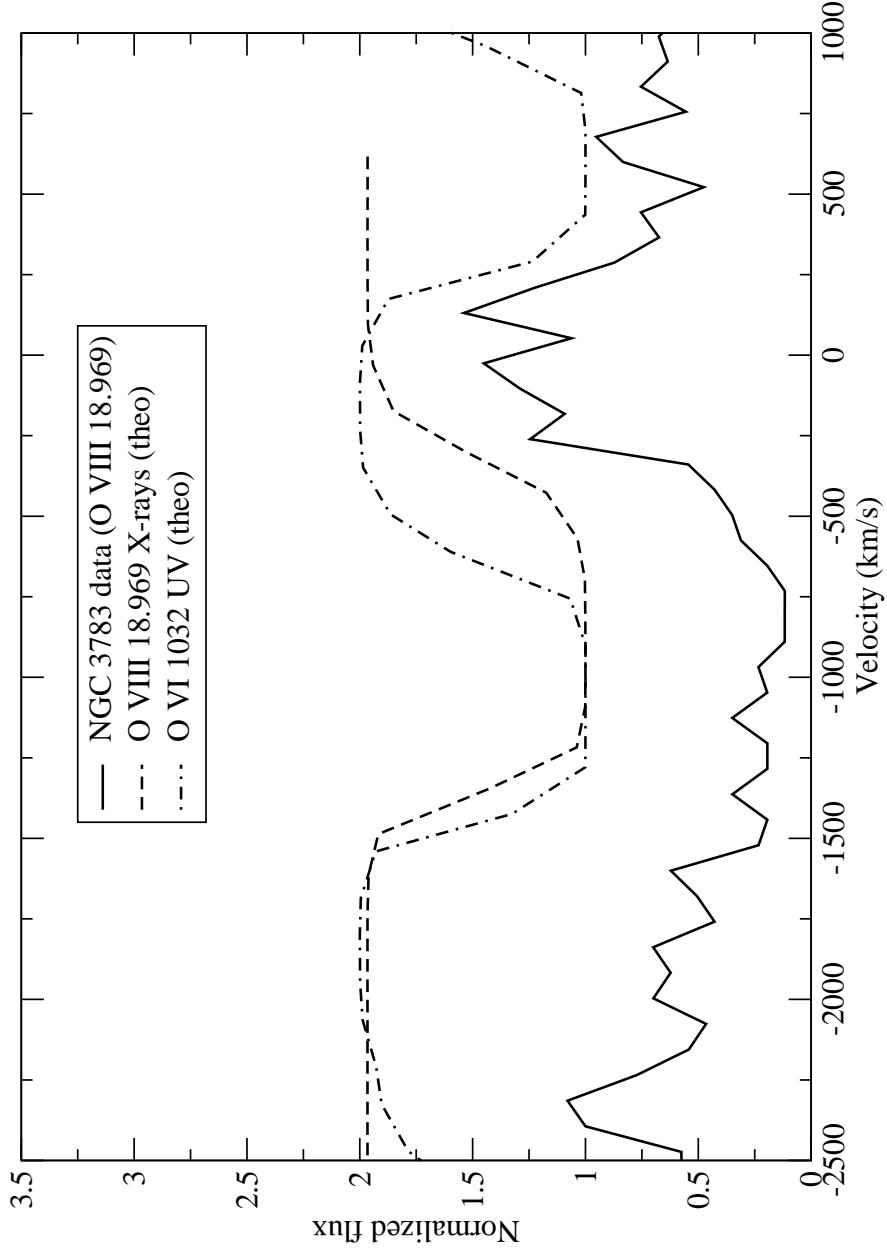}}
\caption{Possible kinematical relationship between the UV and the
X-ray absorbers. In the figure are plotted the theoretical
X-ray and UV lines O~{\sc viii}~$\lambda 18.969$ and 
O~{\sc vi}~$\lambda 1032$, dashed and short-dotted lines
respectively. Also plotted the profile of the line
O~{\sc viii}~$\lambda 18.969$ taken from the 900 ks spectrum of
NGC 3783. All these lines are sharing velocities $\sim$ 1000 \kms.
\label{uv-xrays}}
\end{figure}

The relationship between the X-ray absorption spectrum and the UV absorption 
spectrum is at present a subject of controversy.
\citet{kraemer2001a} and \citet{gabel2003a,gabel2005a} 
have analyzed de UV spectrum
of NGC 3783. HST/STIS and FUSE spectra 
show absorption troughs from the low
order Lyman series (i.e. Ly${\alpha}$,
Ly${\beta}$, Ly${\gamma}$),
C~{\sc iv}~
$\lambda \lambda 1548.2,1550.8$, 
N~{\sc v}~
$\lambda \lambda 1238.8,1242.8$,
O~{\sc vi}~
$\lambda \lambda 1032,1038$. All these lines
are seen in three kinematic components
at $-1365$, $-548$, and $-724$ \kms
(components 1,2 and 3 respectively). A weak
fourth component is reported by \citet{gabel2003a,gabel2005a}
at $-1027$ \kms. Figure \ref{uv700-1700} shows the
spectrum predicted by model B in the range
700--1700 ${\rm \AA}$. It is clear from this
figure that lines due to low order Lyman series,
He~{\sc ii}, C~{\sc iv}, 
N~{\sc v},
O~{\sc vi}, and Ne~{\sc viii} 
would be detectable in the UV
band of the spectrum. Figure \ref{uv-velocity} shows the
spectra of several UV lines as a function of radial
velocity respect to the systemic. At the top of the 
Figure we present the Ly${\alpha}$ blended with
the He~{\sc ii}~$\lambda 1215$ line (upper left) which
yields a feature centered at $\sim -1300$ \kms.
Also an absorption feature is formed with the
Ly${\beta}$ and He~{\sc ii}~$\lambda 1025$ lines (upper center)
with center at $\sim -1200$ \kms, and the
line Ly${\gamma}$ (upper right) with a velocity
$\sim -1100$ \kms.
At the bottom
of the Figure we plot the spectrum of three important 
doublets (C~{\sc iv}~
$\lambda \lambda 1548.2,1550.8$,
N~{\sc v}~
$\lambda \lambda 1238.8,1242.8$,
and O~{\sc vi}~
$\lambda \lambda 1032,1038$). 
The solid line depicts the velocity spectra constructed 
taking the shorter wavelength of the doublet. The spectra
in dashed lines were made  
taking the longer wavelength. One can see
the similarity between the velocities predicted
by our model and the high-velocity components (1, 4 and
likely 3)
seen in the UV spectrum of NGC 3783 and others
Seyfert 1 galaxies (for example NGC~5548).
This is similar to the conclusion reached by \citet{gabel2005a},
based on \citet{kaspi2002a} and \citet{gabel2003a},
where all X-ray lines having sufficiently high-resolution
and S/N were found to span the radial velocities of the
three UV kinematic components (see also next section).
These are predictions which arise naturally from our model
because we are modeling the wind self-consistently with
complete treatment of radiative processes in all wavelengths.
In our models these UV features 
as well as the long wavelength absorption lines
in the X-ray band,
are produced by the low
ionization parameter part of the flow (e.g. $\log \xi = 0.12,-1.13$ 
from Figure \ref{fmodelB} lower right panel).
This demonstrates that the
same absorber can produce
X-ray and UV lines with similar velocities. 
Figure \ref{uv-xrays} shows
the kinematic relationship between the X-ray and 
the UV absorbers. The model has been shifted (up)
for clarity.
Here we show the O~{\sc viii}
$\lambda 18.969$ and the doublet O~{\sc vi}~
$\lambda \lambda 1032,1038$ (with respect to the
shorter wavelength).
For comparison
we plot the histogram data of the
900 ks of NGC 3783. We see that our model is capable of
simultaneously producing X-ray and UV absorption lines with similar
velocities around $\sim 1000$ \kms,
as it has been suggested from UV data \citep{gabel2003a}.

\subsection{Present single wind {\it vs} Multiphase wind scenario}
However, we want to highlight the most important differences
between our model, consisting of a single wind, governed by the laws
of radiative acceleration and the multiphase wind;
subjected to pressure equilibrium.

The first difference is that in our model there is a clear
correlation between ionization state of the ions and
velocity ($\xi-v$), and $\xi$ and number of particle $n_H$ ($\xi-n_H$).
At the same time because of the dependence of $v$ and $n_H$ on the
spatial distance $r$,
a dependence of $\xi$ on $r$.
This is different from the multicomponent in pressure equilibrium
model, suggested by \citet{netzer2003a}.
In that model three components (in ionization)
may coexist in the same volume of space, and lying on the stable
regions of the nearly vertical part of the thermal
stability curve ($logT~vs~log[U/T]$ see Figure 12 in that paper).
Our absorbers are not embedded in an external medium, and they cannot
coexist at exactly the same location but follow the physical laws of radiative
acceleration.
In that context, \citet{gabel2005a} also find that
the UV absorbers of \ngc, may share some properties of the multiphase thermal
wind.
The UV kinematic components 1b, 2 and 3 could occupy the low-temperature
base of the region of the thermal stability curve where a range of temperatures
can coexist at pressure equilibrium.
There is one weak point in the picture of inhomogeneities coexisting at pressure
equilibrium. The low-ionization high-velocity UV absorber does not fit there,
due to a factor of 10 larger pressure than the other component.
If embedded into a more ionized-hotter material, it will
eventually evaporate,
unless there exist an additional confining mechanism.
This extra confining could be provided by magnetic pressure,
requiring moderate magnetic fields ($B\approx 10^{-3}~G$),
as is predicted by some dynamical models
\citep{emmering1992a,deKool1995a}.
In our model, there is no need for confining mechanism,
and if any the more ionized material is closer to the source
shielding the intense continuum radiation preventing
its evaporation.
We conclude by stating that our model is one competitor more,
it cannot be ruled out compared with pressure equilibrium model.
A more detailed comparison between models is beyond the scope of the present
work, that could take place somewhere in the future.

\section{Two outflows}
\label{twoouts1}

So far we have been concerned with the line velocity shifts as measured
with respect to the points of maximum absorption of the lines, but little
have been said about the asymmetry of the troughs.
To quantify the line asymmetries predicted by
the theoretical models we have tabulated the
fraction of the terminal velocity at which
the absorption is maximum, to be compared
with the ratio ($\bar{v}/v_{1}$),
where $v_1$ is the position of the blue edge of each line,
used  to
quantify the asymmetry in the spectrum of \ngc,
see Table 3 of \citet{ramirez2005a}. We characterize
a theoretical line as ``red'' if $v(\tau_{max})/v_{\infty}>0.5$
and as ``blue'' if $v(\tau_{max})/v_{\infty}<0.5$, while for
observed lines they are grouped with respect to
$\bar{v}/v_{1}>0.5$ and
$\bar{v}/v_{1}<0.5$ (see Ramirez et al. 2005).
In Table \ref{tblasim1} we present the observed troughs.
In the first and second columns are the identifications 
and the wavelengths of the lines, in the third
we give the fraction $v(\tau_{max})/v_{\infty}$
for model A, and in the fourth column the
classification from observed lines, as in \citet{ramirez2005a},
i.e., $\bar{v}/v_{1}$.
We classify every trough as either
(R) if it is red or (B) if it is blue. There are clear
discrepancies between
model~A and observation.
In order to improve on the agreement with observations
we found necessary to create a model 
composed by two outflows, which differ
by their $\log \xi_{0}$,
the initial exposition to the source.
In table \ref{twoflows123} we can see the
parameters of the two outflows separately,
which composed model~C. The first flow
with $\log \xi_{0}=3$, which we call
HIF (stand from high ionization flow),
is able to create the majority of the high
ionization lines 
(S~{\sc xvi}~$\lambda$4.729, S~{\sc xv }~$\lambda$5.039,
Fe~{\sc xxiii}~$\lambda$8.303),
with extended blue wings as the flow moves
outwards, and acquires more velocity, in agreement
with the observations. The second, which we call
LIF (stand from low ionization flow), is
characterized by $\log \xi_{0}=2$, is able to create
the extended blue wing of the oxygen lines. This model
has the fundamental ingredients that explain
the asymmetry observed in the lines of NGC 3783 and at
the same time, gives us the bulk velocity of the flow.

In the third column of
Table \ref{tblasim2} we present the characterization 
of the lines
coming from the two-outflows model. The HIF component,
is the major contributor to the formation of the high
ionization lines, i.e. S~{\sc xvi}, Fe~{\sc xxi},
Fe~{\sc xxii}, Fe~{\sc xxiii}, because their optical depths peak
close to $\log \xi_{0}$(HIF), and 
as the velocity
of the flow increases their ionization fraction decrease,
forming their blue wings.
Under this picture, only a few lines are characterized as
blue, while the rest are red. Some intermediate-to-low
ionization ions, have appreciable fractions in the
LIF component, which broadens the lines and yields a red 
apparency to the composed profile.
This effect is illustrated by the fourth column of Table  \ref{tblasim2}.

The composed model reproduces the asymmetry of the lines.
The present uncertainties in the computation of
$v(\tau_{max})$ are of the order of 60 \kms
\space (or $\Delta v(\tau_{max})/v_{\infty}=0.055$
with $v_{\infty}=1100$ \kms). So the lines
classified as pure red  with 
$v(\tau_{max})/v_{\infty}\gtrsim0.55$
and pure blue with 
$v(\tau_{max})/v_{\infty}\lesssim 0.45$
are more reliable than those with
$0.45 \lesssim v(\tau_{max})/v_{\infty}\lesssim 0.55$.

It is interesting to analyze 
why the O~{\sc vii} and O~{\sc viii} lines
are blue, while the lines 
Mg~{\sc xi}~$\lambda$7.473 
and Fe~{\sc xvii}~$\lambda$15.015, remain
red.
The optical depths of Mg~{\sc xi}~$\lambda$7.473,
and Fe~{\sc xvii}~$\lambda$15.015 
are below $10^{-4}$ at velocities
greater than 1100 \kms, so the LIF component
does not contribute to their broadening.
Another interesting example is the difference in classification
between the lines Mg~{\sc xi}~$\lambda$7.473
and Mg~{\sc xi}~$\lambda$9.169. 
The reason is that 
while for Mg~{\sc xi}~$\lambda$7.473 the oscillator
strength is $\sim 5\times 10^{-2}$, the f-value for the line
Mg~{\sc xi}~$\lambda$9.169 is almost one
order of magnitude higher ($\sim 7\times 10^{-1}$), which
makes the optical depth of the latter significant
up to velocities
beyond $\sim 1600$ \kms. 

\begin{figure}
\resizebox{15cm}{!}{\includegraphics[angle=0]{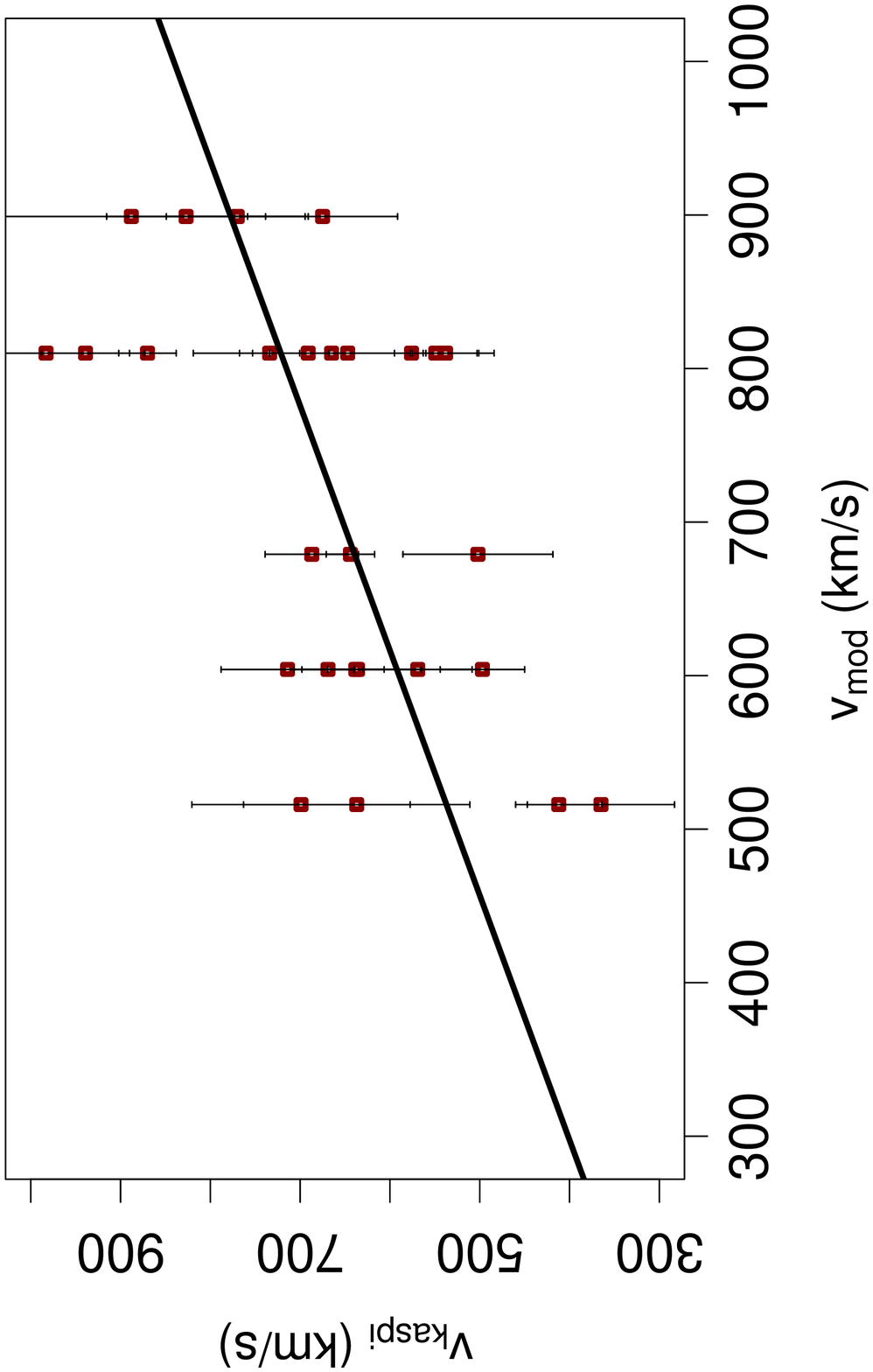}}
\caption{Measured {\it vs} model predicted velocities for
the group of unblended lines given in Table 2 of \citet{ramirez2005a}.
The measurements are taken from \citet{kaspi2002a}.
The solid line is the best line after a linear regression for model C.
\label{mvelC}}
\end{figure}

Figure
\ref{mvelC} compares the modeled
velocities of the lines from the two-outflows model
with observations.
Robust regression using the M estimator
gives a slope of
0.63$\pm$0.22 (solid line in the figure),
and the residual standard error (rms)
is 16.6 for 25 degrees of freedom (dof).

\begin{figure}
\resizebox{15cm}{!}{\includegraphics[angle=0]{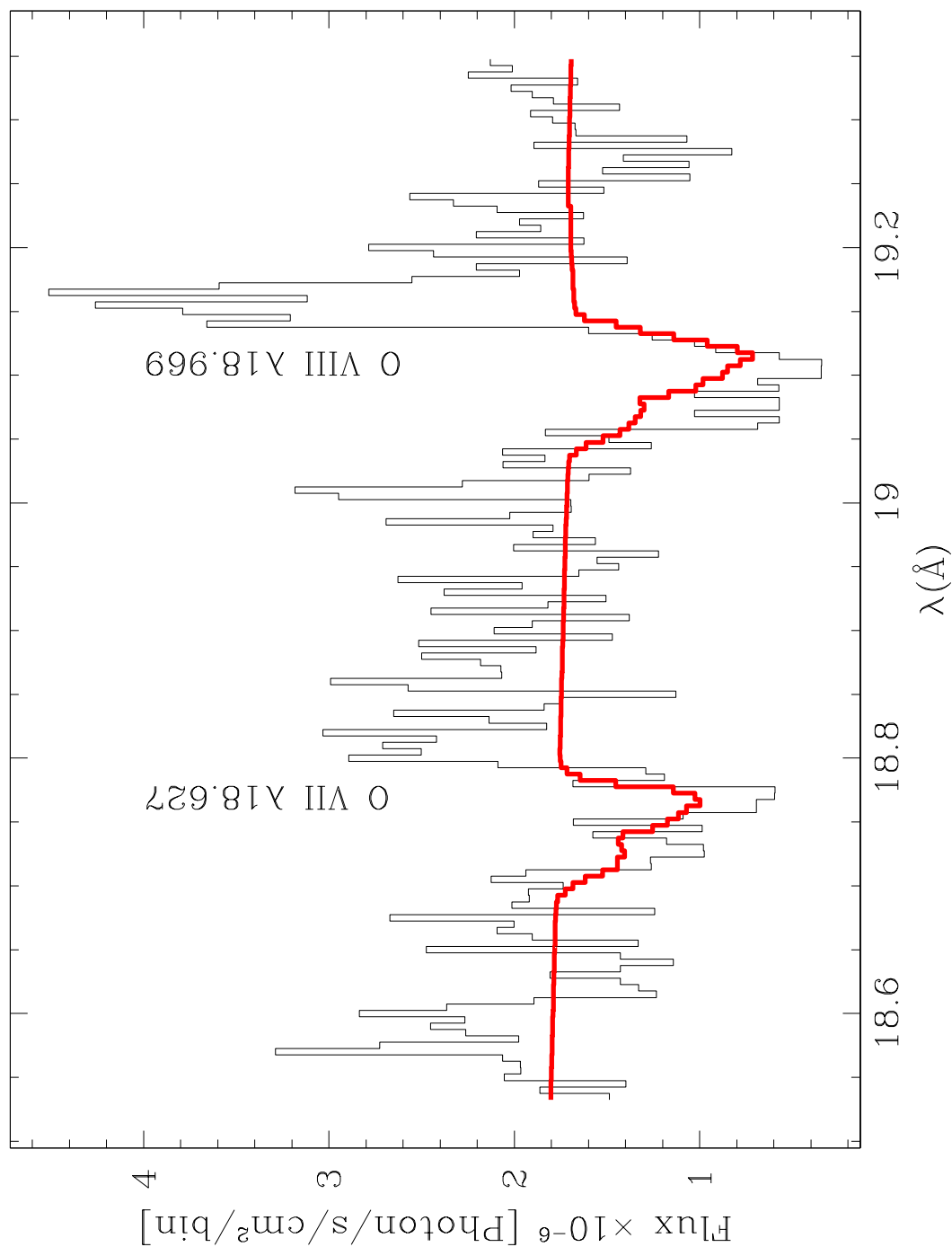}}
\caption{A graphical representation of the flows HIF and LIF
with the line profiles seen in NGC 3783 (histogram). Solid line
is the composed model (model C) by the two flows. It can be seen
that it was necessary to include the second low ionization flow in order
to reproduce the asymmetry of these lines. Otherwise the high
ionization flow would form a more extended red wing.
\label{oviii_modelC}}
\end{figure}

A comparison between the predicted trough profiles and those observed in
NGC~3783 is shown in 
Figure \ref{oviii_modelC}.
Here we plot the theoretical
spectrum given by the outflows HIF and LIF, in the
$18.5-19.3$ ${\rm \AA}$.
Here we have used a powerlaw
as continuum with spectral
index of $\Gamma=1.77$, suitable for this AGN, set
an extra absorber 
{\tt the wabs}
\footnote{N$_{\rm wabs}=0.1 \times 10^{22}$ \cmn.} 
model in XSPEC, 
and the composed fluxes generated by the
HIF and the LIF components.  
To compose the 
O~{\sc viii}~$\lambda$18.969 line profile we have
summed the fluxes from both
flows using the equation (\ref{tautot123}) (where
$m=22$). 
With this approximation the
profile predicted by the two-outflows model is in excellent
agreement with the observation \citet{ramirez2005a}.
The red wing of the troughs, in the range 0--1000 \kms,
is formed by the HIF, while the blue wing, 
in the range $\sim$ 1000--2500 \kms, is
formed by the LIF. 
These values are in agreement
with the values measured for this lines in the
observed spectrum.

It is the first time that such theoretical
work is performed to explain the asymmetry seen
in absorption in the X-ray spectrum of AGNs.
Other works have been able to explain
the blue wings of UV lines, like the well-studied
C~IV $\lambda~1549$,
for example from radiation-driven disk-wind
models \citep{proga2003a}.
So we cannot conclude that the model presented here
is unique for the description
of the asymmetry seen in X-ray lines of AGNs, until
detailed
comparisons are made with such models,
including
X-ray line
profile produced by a wind from a Keplerian accretion
disk
\citep{knigge1995a,shlosman1996a,proga2000a,drew2000a}.
Further work is necessary to be performed for such
comparison and we plan to do that in
a near future.

\section{Summary}
\label{summ1}
We have computed photoionized wind-flow models for the
X-ray spectrum of NGC~3783. We studied singly continously 
absorbing models as well as multiple optically thin components
linked through an analytic wind velocity law.
It is found that the singly continuously absorbing model yields
gas column densities and optical depths too high, unless one adopts very low 
($\sim$ $10^{3-4}$ \cmd) densities and metal abundances ($\sim 0.01$ solar).
On the other hand, the multicomponent
model is able to reproduce observations very well. 
For this model we compute ionization properties of the material using
a velocity law compatible with a radiative wind. 
Our model is consistent  
with $\log n_0 \sim 11.35$ [\cmd],
a launching radii of $\log r_0 \sim 15$ [cm], and a terminal
velocities of $v_{\infty} \sim 1500$ \kms, which yields
a mass loss rate of the order of 
$\dot{M}_{out} \sim 1$ M$_{\sun}$/yr
(assuming a volumic factor $f_{vol}=0.1$).
If we assume an ionizing luminosity of
$L\sim 2 \times 10^{44}$ \ergs
\space \citep{peterson2004a},
and accretion efficiency of $\eta=0.1$,
the Eddington mass accretion mass
is $\dot{M}_{edd} \sim 0.01$ M$_{\sun}$/yr.
This is consistent with the result of \citet{crenshaw2007a} for
NGC~4151,
and \citet{ramirez2008a} for  \apm,
of $\dot{M}_{out}/\dot{M}_{edd} > 10$.
However, it is different from the supposition
made by \citet{goncalves2006a},
of $\dot{M}_{out}/\dot{M}_{edd} \le 1$,
using their photoionization code TITAN,
for computing the single medium in pressure equilibrium.

Finally, the asymmetry seen in the lines of the X-ray
spectrum of NGC 3783 required a model with two
outflows. One flow with a launching ionization
parameter of $\log \xi_0 \sim 3$ and a column
density of $N_H=10^{22}$ \cmn, which recreates 
the red wings of the low ionization lines from
Ne~{\sc ix} to O~{\sc vii}, and the blue wings
of the high ionization lines from Fe~{\sc xxiii}
to Si~{\sc xiv}. A second flow is necessary to
to create the blue wing of the oxygen lines,
which exhibit a blue character, the theoretical
fitting required $\log \xi_0 \sim 2$, terminal
velocities of around $\sim$ 2200 \kms, and a
column density of $N_H=10^{21}$ \cmn.

Our calculations also predict a relationship between the UV and X-ray bands, as
models adjusted to fit the X-ray spectrum naturally predict UV lines 
like the Lyman serie, and the
O~{\sc vi}, N~{\sc v} and C~{\sc iv} doublets,
in apparent concordance with
\citet{costantini2010a}.

\clearpage



\begin{deluxetable}{lc}
\tablecolumns{2}
\tablewidth{0pc}
\tablecaption{Composition and parameter values of the kinematics model A.
\tablenotemark{a}
\label{tbl1solar}}
\tablehead{
\colhead{Element}  & \colhead{Relative Abundance}}
\startdata
H       \nodata & 1.0 \\
He      \nodata & 0.1 \\
C       \nodata & 0.3540E-03 \\
N       \nodata & 0.9330E-04 \\
O       \nodata & 0.7410E-03 \\
Ne      \nodata & 0.1200E-03 \\
Mg      \nodata & 0.3800E-04 \\
Si      \nodata & 0.3550E-04 \\
S       \nodata & 0.2140E-04 \\
Ar      \nodata & 0.3310E-05 \\
Ca      \nodata & 0.2290E-05 \\
Fe      \nodata & 0.3160E-04 \\
\enddata
\tablenotetext{a}{In this model $w_0=0.4$,
$v_{\infty}=900$ ${\rm km}$ ${\rm s}^{-1}$,
$\log _{10} r_0=15.75$ [cm],
$v_{turb}=200$ km~${\rm s}^{-1} $ and
$N_H=5\times 10^{20}$ \cmn .}
\end{deluxetable}


\begin{deluxetable}{lc}
\tablecolumns{2}
\tablewidth{0pc}
\tablecaption{Composition and parameter values of the kinematics model B.
\tablenotemark{a}
\label{tbl2solar}}
\tablehead{
\colhead{Element}  & \colhead{Relative Abundance}}
\startdata
H       \nodata & 1.0 \\
He      \nodata & 0.1 \\
C       \nodata & 0.3540E-03 \\
N       \nodata & 0.9330E-04 \\
O       \nodata & 0.7410E-03 \\
Ne      \nodata & 0.1200E-03 \\
Mg      \nodata & 0.3800E-04 \\
Si      \nodata & 0.3550E-04 \\
S       \nodata & 0.2140E-04 \\
Ar      \nodata & 0.3310E-05 \\
Ca      \nodata & 0.2290E-05 \\
Fe      \nodata & 0.3160E-04 \\
\enddata
\tablenotetext{a}{In this model $w_0=0.4$,
$v_{\infty}=1100$ ${\rm km}$ ${\rm s}^{-1}$,
$\log _{10} r_0=15.75$ [cm],
$v_{turb}=200$ km~${\rm s}^{-1} $ and
$N_H=5\times 10^{20}$ \cmn .}
\end{deluxetable}


\begin{deluxetable}{lrcl} 
\tablecolumns{4}
\tablewidth{0pc} 
\tablecaption{Asymmetry comparison between model A and observation
\label{tblasim1}} 
\tablehead{ 
\colhead{Ion}  & \colhead{Line (${\rm \AA}$)}
& \colhead{$v(\tau_{max})$/900 \kms}
& \colhead{\cite{ramirez2005a}}$^{a}$}
\startdata
 S {\sc xvi}    & $4.729$   & 0.55 (R) & 0.33 $\pm$ 0.1 (B)\\
 S {\sc xv }    & $5.039$   & 0.67 (R) & 0.23 $\pm$ 0.1 (B)\\
 Si {\sc xiii}  & $5.681$   & 0.70 (R) & 0.35 $\pm$ 0.1 (B)\\
 Si {\sc xiv}   & $6.182$   & 0.67 (R) & 0.23 $\pm$ 0.08 (B)\\
 Si {\sc xiii}  & $6.648$   & 0.70 (R) & 0.43 $\pm$ 0.074 (B)\\
 Mg {\sc xii}   & $7.106$   & 0.70 (R) & 0.32 $\pm$ 0.070 (B)\\
 Mg {\sc xi}    & $7.473$   & 0.76 (R) & 0.74 $\pm$ 0.067 (R)\\
 Fe {\sc xxiii} & $8.303$   & 0.55 (R) & 0.037 $\pm$ 0.060 (B)\\
 Mg {\sc xii}   & $8.421$   & 0.70 (R) & 0.32 $\pm$ 0.059 (B)\\
 Mg {\sc xi}    & $9.169$   & 0.76 (R) & 0.31 $\pm$ 0.054 (B)\\
 Ne {\sc x}     & $9.708$   & 0.76 (R) & 0.45 $\pm$ 0.052 (B)\\
 Ne {\sc x}     & $10.240$  & 0.76 (R) & 0.17 $\pm$ 0.048 (B)\\
 Ne {\sc ix}    & $11.547$  & 0.82 (R) & 0.69 $\pm$ 0.043 (R)\\
 Fe {\sc xxii}  & $11.770$  & 0.55 (R) & 0.22 $\pm$ 0.042 (B)\\
 Ne {\sc x}     & $12.134$  & 0.76 (R) & 0.29 $\pm$ 0.041 (B)\\
 Fe {\sc xxi}   & $12.284$  & 0.62 (R) & 0.42 $\pm$ 0.040 (B)\\
 Fe {\sc xviii} & $14.373$  & 0.70 (R) & 0.27 $\pm$ 0.035 (B)\\
 Fe {\sc xviii} & $14.534$  & 0.70 (R) & 0.51 $\pm$ 0.034 (R)\\
 O {\sc viii}   & $14.832$  & 0.82 (R) & 0.21 $\pm$ 0.033 (B)\\
 Fe {\sc xvii}  & $15.015$  & 0.70 (R) & 0.76 $\pm$ 0.033 (R)\\
 O {\sc viii}   & $15.188$  & 0.82 (R) & 0.38 $\pm$ 0.033 (B)\\
 O {\sc viii}   & $16.006$  & 0.82 (R) & 0.26 $\pm$ 0.030 (B)\\
 O {\sc vii}    & $17.200$  & 0.90 (R) & 0.29 $\pm$ 0.029 (B)\\
 O {\sc vii}    & $17.396$  & 0.90 (R) & 0.30 $\pm$ 0.029 (B)\\
 O {\sc vii}    & $17.768$  & 0.90 (R) & 0.32 $\pm$ 0.028 (B)\\
 O {\sc vii}    & $18.627$  & 0.90 (R) & 0.32 $\pm$ 0.027(B)\\
 O {\sc viii}   & $18.969$  & 0.82 (R) & 0.26 $\pm$ 0.026 (B)\\
\enddata

\tablenotetext{a}{$\bar{v}/v_1$, see definition in the text.}

\end{deluxetable}


\begin{deluxetable}{lc} 
\tablecolumns{2} 
\tablewidth{0pc} 
\tablecaption{Parameters of the two outflows model (Model C)
\label{twoflows123}} 
\tablehead{ 
\colhead{Outflow HIF}  & \colhead{Outflow LIF}}
\startdata 
$\log \xi_0=3.0$            & $\log \xi_0=2.0$ \\
$v_{\infty}=1100$ \kms       & $v_{\infty}=2200$ \kms \\
$N_H= 10^{22}$ \cmn          & $N_H= 10^{21}$ \cmn\\
$w_0=0.4$                    & $w_0=0.2$ \\
Solar composition            & Solar composition \\
\enddata
\end{deluxetable}


\begin{deluxetable}{lrcc} 
\tablecolumns{4} 
\tablewidth{0pc} 
\tablecaption{Asymmetry comparison between model C and observation
\label{tblasim2}} 
\tablehead{ 
\colhead{Ion}  & \colhead{Line (${\rm \AA}$)}
& \colhead{$v(\tau_{max}$[HIF]$)$/1100 \kms}
& \colhead{$v(\tau_{max}$[HIF+LIF]$)$/2200 \kms}}
\startdata
 S {\sc xvi}    & $4.729$   & 0.47 (B) & 0.47 (B) \\
 S {\sc xv }    & $5.039$   & 0.55 (R) & 0.55 (R) \\
 Si {\sc xiii}  & $5.681$   & 0.62 (R) & 0.62 (R) \\
 Si {\sc xiv}   & $6.182$   & 0.55 (R) & 0.55 (R) \\
 Si {\sc xiii}  & $6.648$   & 0.62 (R) & 0.62 (R) \\
 Mg {\sc xii}   & $7.106$   & 0.55 (R) & 0.55 (R) \\
 Mg {\sc xi}    & $7.473$   & 0.74 (R) & 0.74 (R) \\
 Fe {\sc xxiii} & $8.303$   & 0.47 (B) & 0.47 (B) \\
 Mg {\sc xii}   & $8.421$   & 0.55 (R) & 0.55 (R) \\
 Mg {\sc xi}    & $9.169$   & 0.74 (R) & 0.37 (B)$^{a}$ \\
 Ne {\sc x}     & $9.708$   & 0.74 (R) & 0.74 (R)  \\
 Ne {\sc x}     & $10.240$  & 0.74 (R) & 0.37 (B)$^{a}$ \\
 Ne {\sc ix}    & $11.547$  & 0.74 (R) & 0.37 (B)$^{a}$ \\
 Fe {\sc xxii}  & $11.770$  & 0.47 (B) & 0.47 (B) \\
 Ne {\sc x}     & $12.134$  & 0.74 (R) & 0.37 (B)$^{a}$ \\
 Fe {\sc xxi}   & $12.284$  & 0.47 (B) & 0.47 (B) \\
 Fe {\sc xviii} & $14.373$  & 0.55 (R) & 0.55 (R) \\
 Fe {\sc xviii} & $14.534$  & 0.55 (R) & 0.55 (R) \\
 O {\sc viii}   & $14.832$  & 0.74 (R) & 0.37 (B)$^{a}$ \\
 Fe {\sc xvii}  & $15.015$  & 0.62 (R) & 0.62 (R) \\
 O {\sc viii}   & $15.188$  & 0.74 (R) & 0.37 (B)$^{a}$ \\
 O {\sc viii}   & $16.006$  & 0.74 (R) & 0.37 (B)$^{a}$ \\
 O {\sc vii}    & $17.200$  & 0.82 (R) & 0.41 (B)$^{a}$ \\
 O {\sc vii}    & $17.396$  & 0.82 (R) & 0.41 (B)$^{a}$ \\
 O {\sc vii}    & $17.768$  & 0.82 (R) & 0.41 (B)$^{a}$ \\
 O {\sc vii}    & $18.627$  & 0.82 (R) & 0.41 (B)$^{a}$ \\
 O {\sc viii}   & $18.969$  & 0.74 (R) & 0.37 (B)$^{a}$ \\
\enddata
\tablenotetext{a}{With the addition of the LIF, we mark those line whose symmetry change.}
\end{deluxetable}


\clearpage


\clearpage

\clearpage

\clearpage

\clearpage

\clearpage


\clearpage

\clearpage

\clearpage

\clearpage

\clearpage

\clearpage

\clearpage

\clearpage

\clearpage

\end{document}